\documentclass[11pt, draftclsnofoot, onecolumn]{IEEEtran}
\usepackage{amssymb}
\usepackage{amsmath}
\usepackage{cite}
\usepackage{url}
\usepackage{xcolor}
\usepackage{cite,graphicx,amsmath,amssymb}
\usepackage{tabularx,colortbl}
\usepackage[english]{babel}
\usepackage{fancyhdr}
\usepackage{subfigure}
\usepackage{citesort}
\usepackage{subfigure}
\usepackage{multirow}
\usepackage{balance}
\usepackage{fancyhdr}
\usepackage{mdwmath}
\usepackage{mdwtab}
\usepackage{caption}
\usepackage{amsthm}
\usepackage[xcolor]{changebar}
\usepackage{lipsum}

\usepackage{flafter}

\newcommand\ytl[2]{
\parbox[b]{8em}{\hfill{\color{cyan}\bfseries\sffamily #1}~$\cdots\cdots$~}\makebox[0pt][c]{$\bullet$}\vrule\quad \parbox[c]{4.5cm}{\vspace{7pt}\color{red!40!black!80}\raggedright\sffamily #2.\\[7pt]}\\[-3pt]}

\newtheorem{theorem}{Theorem}

\newtheorem{lemma}{Lemma}

\newtheorem{corollary}{Corollary}

\makeatletter
\def\ScaleIfNeeded{%
\ifdim\Gin@nat@width>\linewidth \linewidth \else \Gin@nat@width
\fi } \makeatother

\begin{document}

\title{Non-Orthogonal Multiple Access for 5G and Beyond}


\author{


 Yuanwei~Liu,~\IEEEmembership{Member,~IEEE,}
        Zhijin~Qin,~\IEEEmembership{Member,~IEEE,}
        Maged~Elkashlan,~\IEEEmembership{Member,~IEEE,}
         Zhiguo~Ding,~\IEEEmembership{Senior Member,~IEEE,}
         Arumugam~Nallanathan,~\IEEEmembership{Fellow,~IEEE,}
         and Lajos~Hanzo,~\IEEEmembership{Fellow,~IEEE},

\thanks{Y. Liu, M. Elkashlan and A. Nallanathan are with Queen Mary University of London, London,
UK (email: \{yuanwei.liu, maged.elkashlan, a.nallanathan\}@qmul.ac.uk).}
\thanks{Z. Qin and Z. Ding are with Lancaster University, Lancaster, UK (e-mail: \{zhijin.qin,z.ding\}@lancaster.ac.uk).}

 \thanks{L. Hanzo is with University of Southampton, Southampton,
UK (email:lh@ecs.soton.ac.uk).}
}
\maketitle
\begin{abstract}
Driven by the rapid escalation of the wireless  capacity requirements imposed by advanced multimedia applications (e.g., ultra-high-definition video, virtual reality etc.), as well as the dramatically
increasing demand for  user access required for the Internet of Things (IoT), the fifth generation (5G) networks face challenges in terms of supporting large-scale
heterogeneous data traffic. Non-orthogonal multiple access (NOMA), which has been recently proposed for the
3rd generation partnership projects long-term evolution advanced (3GPP-LTE-A), constitutes  a promising technology of addressing the above-mentioned
challenges in 5G networks by accommodating several users within the same orthogonal resource block. By doing so, significant
bandwidth efficiency enhancement can be attained over conventional orthogonal multiple access (OMA) techniques. This motivated
numerous researchers to dedicate substantial research contributions to this field.
In this context, we provide a comprehensive overview of the state-of-the-art in power-domain multiplexing aided NOMA, with a focus
on the theoretical NOMA principles, multiple antenna aided NOMA design, on the interplay between NOMA and cooperative transmission,
on the resource control of NOMA, on the co-existence of NOMA with other emerging potential 5G techniques and on the comparison with other NOMA variants. We highlight the main advantages of power-domain
multiplexing NOMA  compared to other existing NOMA techniques. We summarize the challenges of existing
research contributions of NOMA and provide potential solutions.
Finally, we offer some design guidelines for NOMA systems and identify promising research opportunities for the future.
\end{abstract}


\begin{table}[htbp]\tiny
\caption{LIST OF ACRONYMS}
\begin{center}
\centering
\begin{tabular}{|l|l|}
AF & Amplify-and-Forward\\
BB & Beamforming Based \\
BC & Broadcast Channel\\
BS & Base Station\\
BF & Beamforming\\
CB & Cluster-Based \\
CDMA & Code Division Multiple Access\\
CF & Compress-and-Forward \\
CIRs & Channel Impulse Response \\
CoMP& Coordinated Multipoint \\
CR & Cognitive Radio\\
C-RAN & Cloud-based Radio Access Networks\\
CS & Compressive Sensing\\
CSI & Channel State Information\\
CSIT & Channel State Information at the Transmitter \\
D2D & Device-to-Device\\
DF & Decode-and-Forward\\
DL& Downlink\\
DPC & Dirty-Paper Coding \\
FD & Full-Duplex\\
FDMA & Frequency Division Multiple Access\\
IDMA & Leave Division Multiple Access\\
IMD & Iterative Multi-user Detection\\
IoT & Internet of Things\\
LDS & Low-Density Signature \\
LDPC & Low-Density Parity-Check   \\
LPMA & Lattice Partition Multiple Access\\
LTE & Long Term Evolution\\
LMMSE & Linear Minimum Mean Square Error\\
MA& Multiple Access \\
MAC& Medium Access Control\\
M2M & Machine-to-Machine\\
MNV& Wireless Network Visualization\\
MPA& Message Passing Algorithms \\
MUSA& Multi-User Shared Access \\
MUST& Multi-User Superposition Transmission \\
NP & Non-deterministic Polynomial-time \\
NOMA & Non-Orthogonal Multiple Access\\
OFDM & Orthogonal Frequency Division Multiplexing \\
OFDMA& Orthogonal Frequency Division Multiple Access\\
OMA & Orthogonal Multiple Access\\
PA& Power Allocation\\
PDMA& Pattern Division Multiple Access \\
PF& Proportional Fairness\\
PLS& Physical Layer Secerity\\
P2P & Peer-to-Peer \\
PR& Primary Receiver\\
PT& Primary Transmitter\\
PU& Primary User\\
QoS& Quality of Service\\
RB& Resource Block\\
RF& Radio Frequency\\
RBC& Relaying Broadcast Channel\\
SA& Signal Alignment\\
SC& Superposition Coding\\
SCMA& Sparse Code Multiple Access\\
SDM & Space Division Multiplexing \\
SDMA& Space Division Multiple Access \\
SDN& Software Defined Network\\
SDR& Software Defined Radio\\
SD-NOMA& Software Defined NOMA\\
SIC& Successive Interference Cancelation\\
SISO& Single-Input Single-Output\\
SNR& Signal-Noise Ratio\\
SR& Secondary Receiver\\
ST& Secondary Transmitter\\
SU& Secondary User\\
TCMA& Trellis Coded Multiple Access\\
TDMA& Time Division Multiple Access\\
UL & Uplink\\
V-BLAST& Vertical-Bell Laboratories Layered Space-Time\\
VLC& Visible Light Communication\\
VNI& Visual Network Index\\
VR& Virtual Reality\\
WPT&Wireless Power Transfer\\
ZF & Zero-forcing \\
\end{tabular}
\end{center}
\label{table:abbre}
\end{table}

\section{Introduction}

\subsection{Brief History of Wireless Standardization}
\label{sec:history-wireless-standards}
Following the pioneering contributions of Maxwell and Hertz, Marconi
demonstrated the feasibility of wireless communications across the
Atlantic at the end of the 19th century. By 1928 this technology
became sufficiently mature for the police, the gangsters as well as
for the rich and famous to enjoy tetherless communications.

An early European development was the Swedish Mobile Telephone System
introduced in 1957, which supported 125 users until 1967. In 1966 the
Norwegian system was launched, which operated until 1990.  Following
these, the 1980s led to the roll-out of numerous national mobile phone
systems, most of which relied on analogue frequency modulation and
hence were unable to employ digital error correction
codes. Consequently, their ability to exploit the radical advances in
Digital Signal Processing remained limited.  Hence the achievable
speech quality was typically poor, especially when the Mobile Station
(MS) roamed farther away from the Base Station (BS).

Hence during the 1980s the member states of the European Union launched
a large-scale cooperative research programme, which led to the
standardization of the second generation (2G) system known as the
global system of mobile (GSM) communications.  GSM was the first
digital international mobile system, which rapidly spread across the
globe. The success of GSM shows the sheer power and attraction of
global standardization, motivating competitors to line up behind a
common worldwide solution.

Shortly after the ratification of GSM, a number of other digital
standards emerged, such as the pan-American digital advanced mobile
phone system (D-AMPS) and the direct sequence code division
multiple access (DS-CDMA) based Pan-American system known as IS-95.
IS-95 also had an evolved counterpart, namely the Pan-American
cdma2000 system, which had three parallel CDMA carriers, leading to
the first standardized multi-carrier CDMA (MC-CDMA)
system~\cite{hanzo2003single}.

However, given the consumers' thirst for higher bit rates, during the
early 1990s the research community turned its attention to developing
the third generation (3G) system, which was also based on various CDMA
solutions.  The detailed discussion and the performance
characterization of 3G networks may be found in~\cite{hanzo20083g}.

Despite the 40-year research history of OFDM~\cite{hanzo2005ofdm},
multi-carrier cellular solutions only emerged during the 2000s as the
dominant modulation technique in the context of the 3G partnership
project's (3GPP) long-term evolution (LTE) initiative. Clearly, during
the 2000s multi-carrier solutions have found their way into all the
802.11 wireless standards designed for wireless local area networks
(WLANs), while using different-throughput modem and channel coding
modes, depending on the near-instantaneous channel quality.

What is so beautiful about multi-carrier solutions is their impressive
flexibility, since they have a host of different parameters which
allow us to appropriately configure them and programme them, whatever
the circumstances are - regardless of the propagation environment and
regardless of the quality of service (QoS) requirements, as facilitated by
the employment of adaptive modulation and coding (AMC).
\begin{figure*}[!t]
    \begin{center}
        \includegraphics[width=16cm]{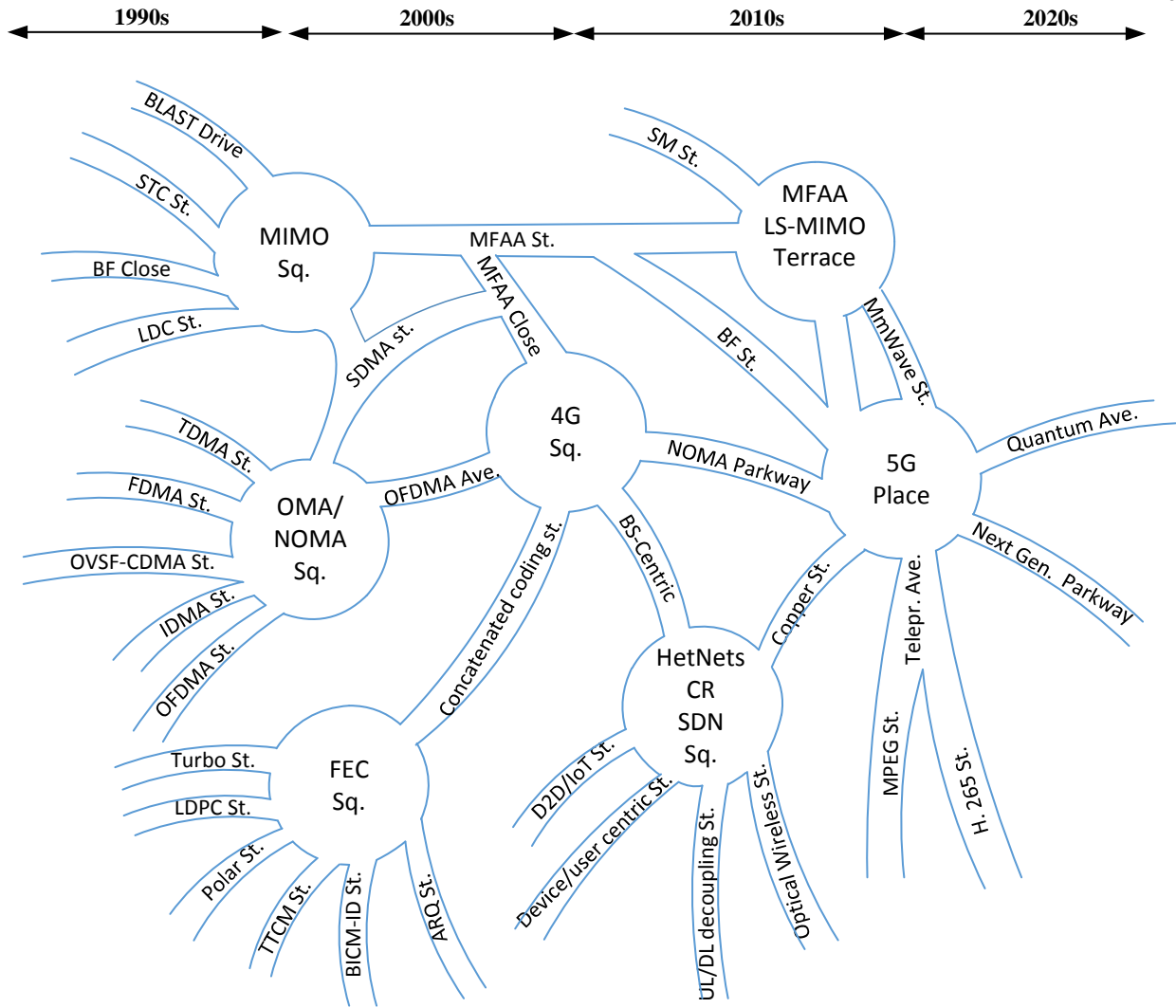}
     \caption{The roadmap for illustrating the brief history of wireless standardization.}
            \label{5G roadmap}
    \end{center}
\end{figure*}


Our hope is dear Colleague that would allow us now to briefly review
the evolution of signal processing and communications techniques over
the past three decades in an anecdotal style with reference to
Fig.~\ref{5G roadmap}. At the time of writing we are gradually approaching
the `5G Place' on our road map of Fig.~\ref{5G roadmap}. We are indeed
also approaching the bit-rate limits upper-bounded by the channel
capacity of both the classic single-input/single-output systems as
well as of the MIMO systems.  Observe at the top left hand corner of
Fig.~\ref{5G roadmap}, how the various MIMO solutions, such as bell lab's
layered space-time (BLAST) `Drive', space-time coding (STC) `Street',
Beam-Forming `Close' and linear dispersion coding (LDC) `Street' merge
into MIMO `Square'.

After decades of evolution, the classic orthogonal multiple access (OMA) schemes, such as time
division multiple access (TDMA) `Street', frequency division multiple
access (FDMA), orthogonal variable spreading factor based code
division multiple access (OVSF-CDMA), interleave division multiple
access (IDMA) and orthogonal frequency division multiple access (OFDMA) `Street' converged to OMA/non-orthogonal multiple access (NOMA) `Square' of
Fig.~\ref{5G roadmap}.  They have also evolved further along spatial
division multiple access (SDMA) and multi-functional antenna array
`Street' - these solutions have found their way into the 4G OFDMA
systems. As seen at the bottom left corner of Fig.~\ref{5G roadmap}, the
various advance channel coding schemes have competed for adoption in
the 4G standard, which relies of a variety of coding arrangements,
including automatic repeat request (ARQ).

At the time of writing the community turned towards the
standardization of the 5G systems, with a special emphasis on the NOMA
techniques detailed in this treatise, as indicated by the broad NOMA
`Parkway', which symbolizes 15 different NOMA proposals.  The family
of MFAAs also entails the recent spatial modulation (SM) and
large-scale (LS) MIMO systems. Since the `road along millimeter wave (mmWave) Street' is
rather unexplored and the attenuation is high, the employment of BF
is rather crucial, if we want to exploit these rich spectral
reserves.

In the bottom right corner of Fig.~\ref{5G roadmap} a number of novel
technological advances converge at HetNet `Square', where cognitive radio (CR) and software defined networks meet device-to-device (D2D)
 and Internet-of-Things (IoT) networks. A range of sophisticated ideas
are also under intensive investigation to resolve the network-centric
versus user-centric design options. There is a strong evidence that the
latter is more promising, because it is also capable of simultaneous
load-balancing. There are also strong proposals on decoupling the
uplink and downlink tele-traffic, with the motivation that
mobile-initiated uplink traffic can reach a small-cell BS at a lower
transmit power than that of the BS's downlink transmission. Optical
wireless based on visible-light communications is also developing
quite rapidly, with Giga-bit copper backhaul networks making promising
progress. Whilst no doubt the classic RF systems will continue to
evolve towards the next generation, an idea, whose time has come is
Quantum communications, as demonstrated by the Science article
``Satellite-based entanglement distribution over 1200 km'' by Yin {\em
  et al.}~\cite{yin2017satellite}.

\begin{figure*}[htbp]
    \begin{center}
        \includegraphics[width=16cm]{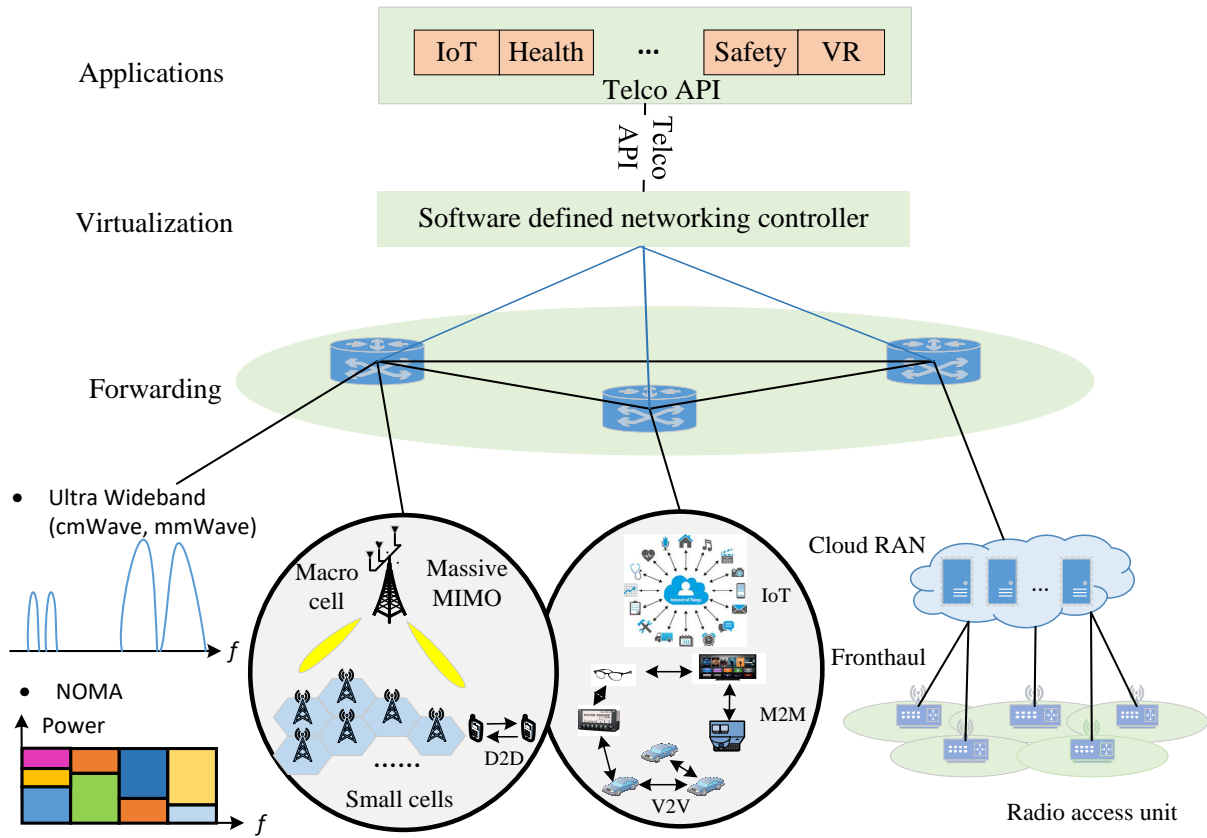}
     \caption{Illustration of the future 5G network architecture.}
            \label{5G architechture}
    \end{center}
\end{figure*}

As the LTE system is reaching maturity and the 4G systems have been commercially deployed, researchers have turned their attention to the 5G cellular network.  The latest visual network index (VNI) reports pointed out that by 2020s, the data traffic of mobile devices will become an order of magnitude higher compared to that in 2014~\cite{CiscoVNI}. Apart from meeting the  escalating data demands of mobile devices, other challenges of dealing with the deluge of data as well as with the high-rate connectivity required by bandwidth-thirsty applications such as virtual reality (VR), online health care and the IoT further aggravate the situation. Driven by this, the 5G networks are anticipated with high expectations in terms of making a substantial breakthrough beyond the previous four generations. The often-quoted albeit potentially unrealistic expectations include 1,000 times higher system capacity, 10 times higher system throughput and 10 times higher energy efficiency per service than those of the fourth generation (4G) networks \cite{wang2014cellular}.
Several key directions such as ultra-densification, mmWave communications, massive MIMO arrangements, D2D and machine-to-machine (M2M) communication, full-duplex (FD) solutions, energy harvesting (EH), cloud-based radio access networks (C-RAN), wireless network virtualization (WNV), and software defined networks (SDN) have been identified by researchers~\cite{boccardi2014five,hossain20155g,andrews2014will}. Fig.~\ref{5G architechture} illustrates the whole 5G network structure, including most of the existing/promising techniques. 

\subsection{State-of-the-art of Multiple Access Techniques}
As mentioned before, sophisticated multiple access (MA) techniques have also been regarded as one of the most fundamental enablers, which have significantly evolved over the consecutive generations in wireless networks \cite{Hanzo2011MIMO,cai2017modulation}. Let us have a deeper looker at the development of MA techniques below. As illustrated in Fig.~\ref{5G roadmap}, the past three to four decades have witnessed historic developments in wireless communications and standardization in terms of MA techniques.  Looking back to the development of the MA formats as we briefly discussed above, in the first generation (1G), FDMA was combined with an analog frequency modulation based technology, although digital control channel signaling was used. In the 2G GSM communications TDMA was used~\cite{steele1999mobile}. Then CDMA, which was originally proposed by Qualcomm \cite{gilhousen1991capacity}, became the dominant MA in the 3G networks. In an effort to overcome the inherent limitation of CDMA - namely that the chip rate has to be much higher than the information data rate - OFDMA was adopted for the 4G networks~\cite{li2013ofdma}. Based on whether the same time or frequency resource can be occupied by more than one user,  the existing MA techniques may be categorized into OMA and NOMA techniques~\cite{Wang2006OMA_NOMA}.  Amongst the above-mentioned MA techniques, FDMA, TDMA and OFDMA allow only a single user to be served within the same time/frequency resource block (RB), which belong to the OMA approach. By contrast, CDMA allows multiple users to be supported by the same RB with the aid of applying different unique, user-specific spreading sequences for distinguishing them.

Fuelled by the unprecedented proliferation of new Internet-enabled
smart devices and innovative applications, the emerging sophisticated
new services expedite the development of 5G networks requiring new
MA techniques.  NOMA techniques can be
primarily classified into a pair of categories, namely,
\emph{code-domain} NOMA and \emph{power-domain}
NOMA~\cite{Dai2015NOMA}\footnote{Note that apart from the code-and
  power-domain, the spatial-domain can also be regarded as another
  domain for supporting multiple users within the same RB, which is
  achieved by exploiting the specific ``spatial signature" constituted
  by the channel impulse responses (CIRs) of the users for
  distinguishing them~\cite{hanzo2005ofdm,Hanzo2011MIMO}. A
  representative MA technique is space division multiple access
  (SDMA)~\cite{SDMA2012Hanzo}.}

The most prominant representatives {\em code-domain NOMA} techniques
include trellis coded multiple access
(TCMA)~\cite{Brannstrom2002TCMA}, IDMA~\cite{liu2006IDMA}, low-density signature (LDS) sequence based
CDMA~\cite{Hoshayar2008LDSCDMA}. These solutions are complemented by
the more recently proposed multi-user shared access (MUSA)
technique~\cite{Yuan2016Multi}, pattern division multiple access
(PDMA)~\cite{ChenPattern7526461}, and sparse code multiple access
(SCMA).

The {\em power-domain NOMA}, which has been recently proposed to
3GPP LTE~\cite{Zhiguo2015Mag}, exhibits a superior capacity region
compared to OMA. The key idea of power-domain NOMA is to ensure that
multiple users can be served within a given time/frequnecy
RB, with the aid of superposition coding (SC) techniques at
the transmitter and successive interference cancellation (SIC) at the
receiver, which is fundamentally different from the classic OMA
techniques of FDMA/TDMA/OFDMA as well as from the code-domain NOMA
techniques.  The motivation behind this approach lies in the fact that
again, NOMA is capable of exploiting the available resources more
efficiently by opportunistically capitalizing on the users' specific channel
conditions~\cite{ding2014pairing} and it is capable of serving
multiple users at different QoS requirements in
the same RB. It has also been pointed out that NOMA has
the potential to be integrated with existing MA paradigms, since it
exploits the new dimension of the power domain.  The milestones of
power-domain NOMA are summarized in the timeline of Table
\ref{table:timeline}.

\begin{table}\tiny
\caption{Timeline of power-domain NOMA milestones.}
\centering
\begin{minipage}[t]{.5\linewidth}
\color{gray}
\rule{\linewidth}{1pt}
\ytl{1972}{Cover first proposed SC and SIC concepts~\cite{cover1972broadcast}}
\ytl{1973}{Bergmans theoretically demonstrated that SC is capable of approaching the capacity of the Gaussian broadcast channel (BC)~\cite{Bergmans1973TIT}}
\ytl{1986}{Verdu discovered the optimal maximum-likelihood multi-user receiver for CDMA systems~\cite{Verdu1986Multiuser}}
\ytl{1994}{Patel and Holtzman proposed to apply SIC and PIC in CDMA systems~\cite{Patel1994CDMA}}
\ytl{2001}{Li and Goldsmith studied the capacity regions for fading BCs with applying SC and SIC~\cite{GoldsmithTIT2001,Goldsmith2TIT2001}}
\ytl{2003}{Mostafa \emph{et al.} demostrated that SAIC can effectively suppress downlink inter-cell interferences in GSM networks\cite{Mostafa2003Mostafa}}
\ytl{2004}{Tse compared the capacity regions of NOMA to OMA both in downlink and uplink~\cite{tse2005fundamentals}}
\ytl{2005}{Andrews summarized the development of interference cancellation for cellular systems~\cite{Andrews2005SIC}}
\ytl{2011}{Zhang and Hanzo offered a unified treatment for SC aided systems~\cite{Zhang2011CST}}
\ytl{2012}{Vanka \emph{et al.} designed an experimental platform for investigating the implementing performance of SC~\cite{Vanka2012SC}}
\ytl{2013}{Saito \emph{et al.} proposed the concept of two-user downlink NOMA transmission for bandwidth efficiency enhancement~\cite{Saito:2013}}
\ytl{2014}{Ding \emph{et al.} developed a multi-user downlink NOMA transmission scheme with randomly deployed users~\cite{ding2014performance}}
\ytl{2015}{Xiong \emph{et al.} designed a practical open source SDR-based NOMA prototype for two-user case~\cite{Xiong2015SDR}}
\ytl{2015}{Benjebbour \emph{et al.} measured the experimental results on a NOMA test-bed for two-user case~\cite{Benjebbour2015testbed}}
\ytl{2015}{Choi \emph{et al.} proposed a two-user MISO-NONA design for investigating the potential application of multi-antenna techniques in NOMA~\cite{Jinho:2015}}
\ytl{2016}{Ding \emph{et al.} proposed a cluster-based multi-user MIMO-NOMA structure~\cite{ding2015mimo}}
\ytl{2017}{Shin \emph{et al.}  proposed to apply coordinated beamforming for multi-cell MIMO-NOMA networks to enhance the cell-edge users' throughput\cite{Shin2017NOMA}}
\bigskip
\rule{\linewidth}{1pt}%
\end{minipage}%
\label{table:timeline}
\end{table}

\subsection{Motivation and Contributions}

While the above literature review has laid the basic foundation for
understanding the development of MA schemes in each
generation of cellular networks, the power domain multiplexing based NOMA
philosophy is far from being fully understood. There are some short magazine papers \cite{Dai2015NOMA,Zhiguo2015Mag,shin2016non,ali2017coordinated} and surveys \cite{Octavia2017survey,tabassum2016non} in the literature that introduce NOMA, but their focus is different from our work. More particularly, Dai \emph{et al.} introduced some concepts of the existing NOMA techniques and identified some challenges and future research opportunities \cite{Dai2015NOMA} both for power-domain and code-domain NOMA. A magazine paper on power-domain NOMA was presented by Ding \emph{et al.} \cite{Zhiguo2015Mag}, with particular attention devoted to investigating the application of NOMA in LTE and 5G networks.  Shin \emph{et al.} \cite{shin2016non} discussed the research challenges and opportunities in terms of NOMA in multi-cell networks, aiming for identifying techniques to manage the multi-cell interference in NOMA. As a further advance, Ali \emph{et al.} \cite{ali2017coordinated} outlined a general framework for multi-cell downlink NOMA by adopting a coordinated multi-point (CoMP) transmission scheme by considering distributed power allocation (PA) strategy in each cell. Regarding surveys, in \cite{Octavia2017survey},  Islam \emph{et al.} have surveyed several recent research contributions on power-domain NOMA, while providing performance comparisons to OMA in different wireless communications scenarios. In \cite{tabassum2016non}, Tabassum \emph{et al.} investigated the uplink and downlink of NOMA in single cell cellular networks, identifying the impact of distance of users on the performance attained.

Although the aforementioned research contributions present either general concepts or specific aspects of NOMA, some important NOMA models, the analytical foundations of NOMA, and some of its significant applications in wireless networks have not been covered. Besides, a clear illustration of the historic development of power-domain NOMA milestones is  missing. Finally, the comparisons between power-domain NOMA and other practical forms of NOMA have not been discussed.  Motivated by all the aforementioned inspirations,  we developed this treatise. More
explicitly, the goal of this survey is to comprehensively survey the state-of-the-art
research contributions that address the major issues, challenges and
opportunities of NOMA, with particular emphasis on both promising new
techniques and novel application scenarios. Table \ref{tab:comparision} illustrates the comparison of this treatise with the existing magazine papers and surveys in the context of NOMA.

\begin{table*}[t!]\tiny 
\begin{center}
{\tabcolsep12pt\begin{tabular}{|l|l|l|l|l|l|l|l|l|l|l|l|}\hline   
  \multirow{3}{*}{---}   & \multirow{3}{*}{Classifications} & \multirow{3}{*}{Key Contents}  & Dai  & Ding  & Ali  & Islam   & Taba.  & Cai & Song & Shin&This  \\
                      & &  & \emph{et al.}   & \emph{et al.}   & \emph{et al.} & \emph{et al.} & \emph{et al.}& \emph{et al.}& \emph{et al.}& \emph{et al.}&work\\

                      &  & & \cite{Dai2015NOMA} & \cite{Zhiguo2015Mag}    &  \cite{ali2017coordinated}  & \cite{Octavia2017survey} & \cite{tabassum2016non}& \cite{cai2017modulation} & \cite{song2016resource} & \cite{shin2016non}&\\
     \hline
\multirow{11}{*}{PD}  & \multirow{2}{*}{Review} &Timeline   & &  & & & && & &$\checkmark$ \\
 \cline{3-12}
                    & & MUD/IC &  &   &  & & & & & & $\checkmark$\\

\cline{2-12}
                    & \multirow{2}{*}{Basics}  & Uplink NOMA & $\checkmark$  & $\checkmark$  &   & $\checkmark$ & $\checkmark$ & $\checkmark$ & $\checkmark$ & $\checkmark$& $\checkmark$\\
   \cline{3-12}
                    & & Fairness &  &  $\checkmark$ &    &$\checkmark$ &    & $\checkmark$& &$\checkmark$ & $\checkmark$\\
 \cline{2-12}
                    & Performance & Cooperative  &  &  $\checkmark$  &    & $\checkmark$ & $\checkmark$ & $\checkmark$ & $\checkmark$& &$\checkmark$\\
 \cline{3-12}
                   &  Enhancement & MIMO-NOMA  & $\checkmark$ & $\checkmark$  &    & $\checkmark$& $\checkmark$& $\checkmark$& & &$\checkmark$\\
 \cline{3-12}
                   &  & Multi-cell CoMP & $\checkmark$ &   & $\checkmark$ & $\checkmark$ & $\checkmark$& & & $\checkmark$&$\checkmark$\\
 \cline{2-12}
                    & Resource & User Scheduling & $\checkmark$ &  $\checkmark$ & $\checkmark$ & $\checkmark$ & $\checkmark$ & $\checkmark$& $\checkmark$& &$\checkmark$\\
  \cline{3-12}
                    & Management & Power Allocation &  &   & $\checkmark$ & $\checkmark$& $\checkmark$& $\checkmark$& $\checkmark$& $\checkmark$&$\checkmark$\\

  \cline{3-12}
                    &  & SD-NOMA & $\checkmark$ &   &  & & & & & &$\checkmark$\\
 \cline{2-12}
                     & Practical  & EP of SIC&    &     &     & $\checkmark$ & $\checkmark$&        &     & $\checkmark$&$\checkmark$\\
   \cline{3-12}
                   & Issues & CE Errors& $\checkmark$ & $\checkmark$  &  &$\checkmark$ &     &     &    &   $\checkmark$  & $\checkmark$\\
      \cline{3-12}
                   &  & Quantization  &  &   &  &$\checkmark$ & & & & & $\checkmark$\\
         \cline{3-12}
             &  & Synchronization &  &   &  & $\checkmark$& & & & & $\checkmark$\\
         \cline{3-12}
                   &  & Security &  &   &  & & & & &  $\checkmark$& $\checkmark$\\
     \cline{2-12}
                   & \multirow{2}{*}{Compatibility}  & HetNets-NOMA &  &   & $\checkmark$ & $\checkmark$   &    & $\checkmark$& & &$\checkmark$\\
      \cline{3-12}
       &   & MmWave-NOMA &  &   &  & & & & & &$\checkmark$\\
             \cline{3-12}
       &   & CR-NOMA &  & $\checkmark$  &  & & & & & &$\checkmark$\\
                    \cline{3-12}
       &   & D2D-NOMA &  &   &  & & & & & &$\checkmark$\\
      \cline{3-12}
                         &  & Modulation &  &   &  & & & $\checkmark$& & &\\
      \cline{2-12}
                    & Progress & Standardization & $\checkmark$ &     & $\checkmark$ & & & & & & $\checkmark$\\

\hline
\multirow{7}{*}{CD} &  \multirow{4}{*}{Single-Carrier}  & IDMA &  & &   & & & & & &$\checkmark$\\
\cline{3-12}
              &  & LDS-CDMA  & $\checkmark$ &   &    & & & $\checkmark$& & &$\checkmark$\\
\cline{3-12}
              &  &LPMA  &   &    &   & &  & $\checkmark$& & &$\checkmark$\\
   \cline{3-12}
              &  & MUST &   & $\checkmark$  & & & & & & &\\
\cline{2-12}

              & \multirow{4}{*}{Multi-Carrier}  & LDS-OFDM  &$\checkmark$ &     &   & &  &  $\checkmark$& & &$\checkmark$\\
 \cline{3-12}
              &  & SCMA & $\checkmark$  &   & & & &$\checkmark$ & & &$\checkmark$\\
  \cline{3-12}
              &  & PDMA &   &   & & & &  $\checkmark$& & &$\checkmark$\\
   \cline{3-12}
              &  & MUSA & $\checkmark$  &   & & & & & & &\\
\hline

\end{tabular}}
\end{center}
\caption{Comparison of this work with available magazine papers and surveys. Here, ``PD" refers to ``Power-Domain", ``CD" refers to ``Code-Domain", ``SD" refers to ``Software-Defined", ``EP" refers to ``Error Propagation", and ``CE" refers to ``Channel Estimation". }
\label{tab:comparision}
\end{table*}

To highlight the significance of this contribution, we commence with a
survey of NOMA starting with the basic principles, which provides the
readers with the basic concepts of NOMA. We continue in the context of
multiple antenna aided techniques combined with NOMA, followed by
cooperative NOMA techniques. We then address another important issue
of NOMA, namely its resource and PA problems. Finally, we
elaborate on invoking other 5G candidate techniques in the context of NOMA
networks. The contributions of this survey are at least five fold,
which are summarized as follows:

\begin{enumerate}
    \item We present a comprehensive survey on the recent advances and
      on the state-of-the-art in power-domain multiplexing aided NOMA
      techniques. The basic concepts of NOMA are introduced and key
      advantages are summarized.  The research challenges,
      opportunities and potential solutions are also identified.
    \item We investigate the application of multiple-antenna aided
      techniques to NOMA. The pair of most dominant solutions - namely
      cluster-based MIMO-NOMA and beamformer-based MIMO-NOMA - are
      reviewed and their benefits are examined. Furthermore, we
      highlight that specific massive-MIMO-NOMA solutions are capable
      of improving the performance of NOMA networks to a large
      extent. A range of important challenges are elaborated on in the
      context of multiple-antenna assisted NOMA and the associated
      future opportunities are also underlined.
    \item By exploiting the specific characteristics of NOMA, we study
      the interplay between NOMA and cooperative communications. We
      demonstrate that cooperative NOMA constitutes a promising
      technique of improving the reliability of the users experiencing
      poor channel conditions.
     \item We identify the potential issues associated both with
       power- and user-allocation, which constitute the fundamental
       problems to be solved for ensuring fairness in the NOMA
       networks. We point out the significance of designing efficient
       algorithms for dynamically allocating the resources to the
       users. Furthermore, we propose the novel concept of a
       software-defined NOMA (SD-NOMA) network architectures, where
       resource allocation - including the power - is performed on a
       generic hardware platform by taking into account the global
       view of the entire network.
     \item We identify the major issues and challenges associated with
       the co-existence of NOMA and the other emerging 5G
       technologies. The potential solutions based on the current
       research contributions corresponding to these technologies are
       also surveyed. We have also discussed the implementation issues and recent standardization progress for NOMA. Finally, power-domain NOMA and other popular forms of NOMA are contrasted. We spotlight that the  significance to provide a unified framework for NOMA.
\end{enumerate}

\subsection{Organization}

\begin{figure*}[t!]
    \begin{center}
        \includegraphics[width=12cm]{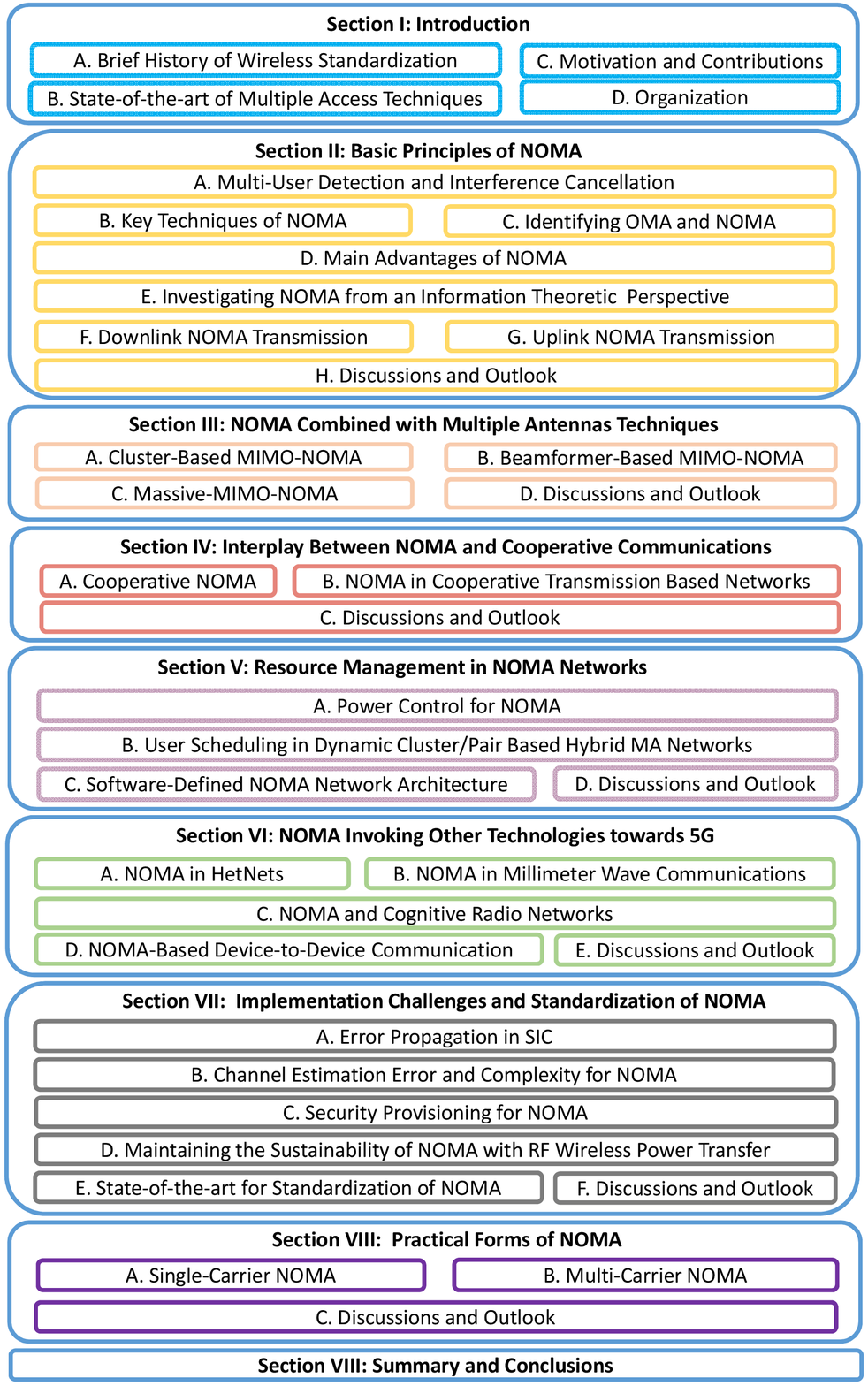}
     \caption{Condensed overview of this survey on NOMA}.
            \label{NOMA structure}
    \end{center}
\end{figure*}

The remainder of this paper is structured as follows. Section II
presents the basic principles of NOMA, including a brief overview on multi-user detection and interference cancellation (IC), the key techniques
adopted and its main advantages. Section III investigates the recent
advances in multiple-antenna aided NOMA transmission and identifies
the open research challenges. In Section IV, the associated research
contributions are surveyed in terms of the interplay between NOMA and
cooperative transmission. The resource allocation of NOMA - including
user association and PA - are studied and summarized in
Section V. Finally, Section VI investigates the co-existence of NOMA
with other emerging 5G technologies. Section VII points out the implementation challenges as well as the standardization process of NOMA, while Section VIII discusses several other forms of NOMA techniques. Finally, Section IX concludes this
treatise. Fig. \ref{NOMA structure} illustrates the organization of this
paper.


\section{Basic Principles of NOMA}
The fundamental concept of NOMA facilitates supporting multiple
users in the power domain. In
contrast to the conventional multiple access (MA) techniques, NOMA\footnote{The NOMA approach
  addressed in this survey implies the power-domain NOMA,
  unless stated otherwise.}
uses a new dimension to perform multiplexing within one of the classic
time/frequecy domains. In other words, NOMA can be regarded as an
``add-on" technique, which has the promising potential of facilitating
harmonious integration with the existing legacy-solutions. In this section,
we introduce the basic concept of NOMA by illustrating the associated key
technologies and summarize its salient advantages.  We also survey the
pivotal research contributions both in downlink and uplink NOMA
transmissions in single-antenna scenarios.

\subsection{Muti-User Detection and Interference Cancelation}
Before diving into the deep  description of the key techniques of NOMA, let us first give a brief overview for the multi-user detection (MUD) and IC.

\subsubsection{Muti-User Detection}
The idea of simultaneously receiving the multiple users signals  and detecting them is not new, its history goes back to the 1970s, please see Cover's work on BCs \cite{cover1972broadcast} and Etten's work on multi-user sequence estimator \cite{Etten1976ML}. Then Verdu discovered the optimal multi-user receiver of CDMA systems in the 1980s \cite{Verdu1986Multiuser}, which is based on the maximum-likelihood receiver detecting multiple users in parallel. We take CDMA as an example to illustrate MUD.
Rather than allocating orthogonal RBs to different users as in FDMA and TDMA, CDMA codes superimposes the signals of multiple users with the aid of unique, user-specific signature referred to as spreading codes. If we use orthogonal $K$-chip, we can only support $K$ users. But when using non-orthogonal $m$-sequences in the spirit of NOMA, we can have many more users. Multiple access interference (MAI) limits the capacity and performance of CDMA systems. While the MAI caused by a single interfering user is generally small, the system becomes interference-limited as the number of interferers or their power increases~\cite{CDMA:PM:1995,CDMA:CM:1996}.  MUD exploiting the knowledge of both the spreading code and timing (and possibly amplitude and phase) information of multiple users has been regarded as an efficient strategy of improving the system capacity. Various MUD algorithms, such as the optimal maximum-likelihood sequence estimation~\cite{Verdu1986Multiuser,MLSE:1991}, turbo decoding,  matched filter SIC~\cite{SIC1990}, and parallel interference cancellation (PIC) have been designed to reduce the MAI at an affordable complexity cost. Table 1 of \cite{Andrews2005SIC} illustrates the high-level comparison for different types of multi-user receivers.


Moreover, there are some recently developed joint detection techniques for downlink cellular systems, which are based on single-antenna interference cancellation (SAIC) receivers. More particularly, this technique relies on either maximum-likelihood detection or predetection processing, rather than on IC techniques. This development is attributed to the fact that joint detection has also been developed for asynchronous networks \cite{Schoeneich2003SAIC}. As a further advance, the proposed SAIC technique has had successful field trial results in the GSM era in terms of suppressing the downlink inter-cell interference \cite{Mostafa2003Mostafa}.

\subsubsection{Interference Cancellation}

Commonly, the multi-user IC  techniques can divided into two main categories, namely pre-interference cancelation (pre-IC) and post-interference cancellation (post-IC) \cite{Miridakis2013SIC}. More specifically, pre-IC techniques are employed at the transmitter side by suppressing the interference by precoding approaches, such as the famous dirty paper coding (DPC) \cite{Costa1983DPC} upon exploiting the knowledge of  the channel state information (CSI) at the transmitter.  By contrast, the post-IC techniques are usually used at the receiver side for cancelling the interference. The post-IC approach can be further divided into two categories, which are parallel and successive \cite{Andrews2005SIC}. If we carry out accurate power control to ensure that all received signals are similar, PIC outperforms SIC. By contrast, SIC works better, when the received powers are different \cite{Patel1994CDMA} because the strongest user's signal can be detected first. The detected bit is the remodulate and its interference is deducted from the received signal. Repeating this  action in a sequential order gives us the clean weakest signal. It is worth noting that in addition to the performance versus complexity trade-off, there are also a variety of other trade-offs between PIC and SIC.

There are also some recent IC techniques. Based on PIC, Sawahashi \emph{et al.} \cite{Sawahashi2002CDMA} of NTT DoCoMo have developed a  coherent three-stage interference canceller for DS-CDMA systems, which is capable of employing accurate pilot symbol-assisted channel estimation and IC. The corresponding performance was experimentally evaluated with the aid of a multipath fading simulator \cite{Sawahashi2002CDMA}. Regarding the LTE Heterogeneous networks (HetNets), some companies such as Qualcomm and Fujitsu have also developed powerful IC techniques. More particularly, in the 3GPP  meeting in 2008, Qualcomm proposed a new cell selection scheme, termed cell range extension \cite{3GPP2008Qualcomm}, for allowing more small cell offloading to manage the strong macrocell interference at the handsets \cite{Andrews2014loadlalancing}.  Li \emph{et al.} \cite{Li2013LTE} of Fujitsu proposed a frequency domain  cell-specific reference signal aided SIC scheme for 3GPP LTE-Advanced Rel-11 systems. While the aforementioned IC techniques mainly focus on the digital domain, there are also pioneering analog domain solutions. In full-duplex systems, self interference cancellation (IC) is the key issue \cite{Duarte2012fullduplex}. Duarte \emph{et al.} \cite{Duarte2010} proposed an active analog cancellation aided full-duplex architecture to cancel the self-interference before the analog-to-digital converter (ADC). The suppression performance of \cite{Duarte2010} was also characterized with the aid of experimental results.

\subsection{Key Technologies of NOMA}
 Again, the basic principles of NOMA techniques rely on the employment
 of superposition coding (SC) at the transmitter and successive
 interference cancelation (SIC) techniques at the receiver. In fact, neither of
 these two techniques are new, their roots can be found in the
 existing
 literature~\cite{cover1972broadcast,cover2012elements,Bergmans1973TIT,Carleial,Cover1979relay,Grant2001MA,Csiszar1978wiretap,Vanka2012SC,Andrews2005SIC,Patel1994CDMA,Wolniansky1998VBLAST,gelal2013topology,Jiang2012SIC,Xu2013SIC,Lee2013SIC}.

\subsubsection{Superposition Coding}
First proposed by Cover as early as in 1972~\cite{cover1972broadcast},
the elegant idea of SC is regarded as one of the fundamental building
blocks of coding schemes conceived for achieving the capacity of a
scalar Gaussian BC~\cite{cover2012elements}. More
particularly, it was theoretically demonstrated that SC is capable of
approaching the capacities of both the Gaussian BC and
of the general BC as defined by Bergmans' paper
published as early as 1973~\cite{Bergmans1973TIT} and by Gallager
\cite{gallager1974capacity}. The fundamental concept of SC is that it
is capable of encoding a message for a user associated with poor
channel conditions at a lower rate and then superimpose the signal of
a user having better channel conditions on it. Inspired by the solid
foundations laid down from an information theory perspective,
researchers became motivated to apply SC to diverse channels, such as
interference channels~\cite{Carleial}, relay channels
\cite{Cover1979relay}, MA channels~\cite{Grant2001MA}
and wiretap channels~\cite{Csiszar1978wiretap}. While the
aforementioned contributions richly motivate the use of SC from a
theoretical perspective, further research was required for evolving
this technique from theory to
practice~\cite{Hanzo2011SC,Vanka2012SC}. Specifically, Vanka \emph{et
  al.}~\cite{Vanka2012SC} designed an experimental platform using a
software defined radio (SDR) system to investigate the performance of
SC. The set of achievable rate-pairs under a specific packet-error
constraint was determined.

\subsubsection{Successive Interference Cancelation}
It has been widely exploited that the network capacity can be
substantially improved with the aid of efficient interference
management, hence SIC is regarded as a promising IC technique in wireless networks.  By invoking the
following procedure, it enables the user having the strongest signal
to be detected first, who has hence the least
interference-contaminated signal. Then, the strongest user re-encodes
and remodulates its signal, which is then subtracted from the
composite signal. The same procedure is followed by the second
strongest signal, which has in fact become the strongest signal.  When
all but one of the signals was detected, the weakest user decodes its
information without suffering from any interference at all.

Compared to the classic MUD techniques, we summarize the key advantages of SIC as follows:
\begin{itemize}
  \item In contrast to the MUD techniques of CDMA \cite{hanzo2007ofdm} or MIMO systems \cite{hanzo2010mimo} where there are multiple observations at receivers \cite{Verdu1986Multiuser,Hallen1995CDMA,Moshavi1996DSCDMA}, power-domain NOMA usually has a single observation at each receiver. In other words, rather than using join detection aided MUD considering all users simultaneously, SIC technique operate in an iterative manner, which imposes a lower hardware complexity at the receiver than the joint decoding approach~\cite{Andrews2005SIC}.\footnote{Actually, if we consider the broad sense of power-domain NOMA, most of aforementioned MUD techniques can be potentially applied, depending on the receiving power of users.} The number of multiplications and additions required was summarized in Hanzo \emph{et al} \cite{hanzo2007ofdm}.
  \item It has been demonstrated that SIC is capable of approaching the boundaries of the capacity region of both the BCs and of MA channels~\cite{tse2005fundamentals}.
  \item As discussed before, SIC has a better performance than PIC when the received powers are different \cite{Patel1994CDMA}, which is more suitable for power-domain NOMA supporting the users via different power levels.
\end{itemize}

Hence,
SIC has been widely studied and diverse versions have been employed in
practical systems, such as CDMA~\cite{Patel1994CDMA} as well as SDMA \cite{hanzo2007ofdm} and the
vertical-bell laboratories layered space-time
(V-BLAST)~\cite{Wolniansky1998VBLAST}. Furthermore, SIC has also been
exploited in several practical scenarios, such as multi-user MIMO
networks~\cite{gelal2013topology}, multi-hop
networks~\cite{Jiang2012SIC}, random access systems~\cite{Xu2013SIC},
and in large-scale networks modeled by stochastic
geometry~\cite{Lee2013SIC}, just to name a few.  A particularly important fact concerning
SIC is that it has been implemented in commercialized systems, such as
IEEE 802.15.4.  Another practical implementation is to use the
SIC-aided spatial division multiplexing (SDM) detector in SDM-assisted
orthogonal frequency division multiplexing (OFDM) systems for signal
detection by distinguishing the users with the aid of their unique,
user-specific impulse responses~\cite{hanzo2005ofdm}.

\subsection{Identifying OMA and NOMA}

We commence with the mathematical definition of NOMA. \emph{The general definition of NOMA is a MA technique, which allows multiple users to simultaneously occupy the same time-and-frequency resource}. Based on this definition, we may have power-domain NOMA, code-domain NOMA and spatial-domain NOMA, as mentioned in Section I. In this treatise, we focus on power-domain NOMA. \emph{The narrow sense definition of power-domain NOMA is to superimpose the signals in the same time-and-frequency resource at different power levels, and then to adopt SIC techniques at receivers for interference cancelation}.\footnote{The broad sense definition of power-domain NOMA constitutes all MA techniques which apply SC technique at transmitters for power multiplexing. In this sense, some code-domain NOMA can also regarded as generalized power-domain NOMA techniques, which will be detailed in Section VII.}

For illustrating the relationship between NOMA and OMA mathematically, below we provide a simple analytical characterization by examining the achievable  performance with the aid of signal-noise ratio (SNR) expressions. Let us consider two-user downlink NOMA transmission. The channel coefficients of user $m$ and user $n$ are $h_m$ and $h_n$. Let us denote the transmit SNR at the BS by $\rho$ and assume that we have $|h_m|^2<|h_n|^2$ without loss of generality.
\begin{itemize}
\item \textbf{OMA}: According to Shannon's capacity theorem, when power control is used, the throughput of OMA  can be expressed for user $m$ and user $n$ as  \cite{tse2005fundamentals}
\begin{align}\label{OMA_m power control}
R_m^{\mathrm{OMA}} = \beta {\log _2}\left( {1 + \frac{{{\alpha _m}\rho }}{\beta }|{h_m}{|^2}} \right),
\end{align}
and
\begin{align}\label{OMA_n power control}
R_n^{\mathrm{OMA}} = (1 - \beta ){\log _2}\left( {1 + \frac{{{\alpha _n}\rho }}{{1 - \beta }}|{h_n}{|^2}} \right),
\end{align}
respectively, where $\alpha_m$ and $\alpha_n$ are the PA coefficients and satisfy $\alpha_m+\alpha_n=1$, and $\beta$ is the resource allocation coefficient, having units of `Hz' for frequency or `s' for time.  For the case that the power control is not considered at the BS, we set the $\frac{{{\alpha _m}}}{\beta } = \frac{{{\alpha _n}}}{{1 - \beta }} = 1$. Then  \eqref{OMA_m power control} and \eqref{OMA_n power control} can be rewritten as
\begin{align}\label{OMA_m}
R_m^{\mathrm{OMA}}= \beta {\log _2}\left( {1 + \rho |{h_m}{|^2}} \right),
\end{align}
and
\begin{align}\label{OMA_n}
R_n^{\mathrm{OMA}} = (1 - \beta ){\log _2}\left( {1 + \rho |{h_n}{|^2}} \right).
\end{align}

\item \textbf{NOMA}: Regarding NOMA, the throughput of the user $m$ and user $n$  is given by \cite{tse2005fundamentals}
  \begin{align}\label{NOMA_m}
R_m^{\mathrm{NOMA}} = {\log _2}\left( {1 + \frac{{\rho {\alpha _m}|{h_m}{|^2}}}{{1 + \rho {\alpha _n}|{h_m}{|^2}}}} \right),
\end{align}
and
  \begin{align}\label{NOMA_n}
R_n^{\mathrm{NOMA}} = {\log _2}\left( {1 + \rho {\alpha _n}|{h_n}{|^2}} \right),
\end{align}
respectively.
\end{itemize}

In order to gain more insights into the spectral efficiency advantage of NOMA over OMA, we investigate the following special case as an example. At high SNRs, assuming that the time/frequency resources are equally allocated to each user, based on \eqref{OMA_m}, \eqref{OMA_n}, \eqref{NOMA_m}, and \eqref{NOMA_n}, the sum throughput of OMA and NOMA can be expressed as $R_{sum,\infty }^{OMA} \approx  {\log _2}\left( {\rho \sqrt {|{h_m}{|^2}|{h_n}{|^2}} } \right)$ and  $R_{sum,\infty }^{NOMA} \approx {\log _2}\left( {\rho |{h_n}{|^2}} \right)$, respectively. Then we can express the sum throughput gain of NOMA over OMA as follows:
\begin{align}\label{NOMA gain}
R_{sum,\infty }^{\mathrm{Gain}} = R_{sum,\infty }^{\mathrm{NOMA}} - R_{sum,\infty }^{\mathrm{OMA}} = \frac{1}{2}{\log _2}\left( {|{h_n}{|^2}/|{h_m}{|^2}} \right).
\end{align}

When we have $|h_m|^2<|h_n|^2$,  the sum throughput of NOMA is higher than that of OMA, and this gain is imposed  when the channel conditions of the two users become more different. Chen \emph{et al.}~\cite{chen2016mathematical} provided a rigorous mathematical proof to demonstrate that NOMA always outperforms the conventional OMA scheme.

\subsection{Main Advantages of NOMA}

Whilst both SC and SIC continue to mature in terms of their
theoretical and practical aspects, NOMA is also maturing. Hence it has
been proposed for next-generation networks. By invoking the SC
technique, the BS transmits the superposition coded signals of all
users. Then the channel gains of the users are sorted in the
increasing or decreasing order. In the traditional OMA schemes, strongest user benefits from the highest power, which is not always the case for NOMA.  The NOMA transmission schemes exhibit the
following main advantages.


\begin{itemize}
  \item \textbf{High bandwidth efficiency:} NOMA exhibits a high
    bandwidth efficiency and hence improves the system's throughput,
    which is attributed to the fact that NOMA allows each RB (e.g., time/frequecy) to be exploited by multiple
    users~\cite{Saito:2013}.
  \item \textbf{Fairness}: A key feature of NOMA is that it allocates
    more power to the weak users.  By doing so, NOMA is
    capable of guaranteeing an attractive tradeoff between the
    fairness among users in terms of their throughput. There are
    sophisticated techniques of maintaining fairness for NOMA, such as
    the intelligent PA policies
    of~\cite{Timotheou:2015,ding2014pairing} and the cooperative NOMA
    scheme of~\cite{Ding2015cooperative}, which will be detailed in
    this paper later.
  \item \textbf{Ultra-high Connectivity}: The future 5G system is
    expected to support the connection of billions of smart devices in
    the IoT \cite{shirvanimoghaddam2016massive}.  The existence of NOMA offers a promising design
    alternative for efficiently solving this non-trivial task by
    exploiting its non-orthogonal characteristics.  More specifically,
    in contrast to convectional OMA, which requires the same number of frequency/time
    RBs as the number of devices, NOMA is capable of serving them by
    using less RBs.
  \item \textbf{Compatibility}: Again, from a theoretical perspective,
    NOMA can be invoked as an ``add-on" technique for any existing OMA
    techniques, such as TDMA/FDMA/CDMA/OFDMA, due to the fact that it
    exploits a new dimension, namely the power-domain. Given the
    mature status of SC and SIC techniques both in theory and
    practice, NOMA may be amalgamated with the existing MA techniques.
   \item \textbf{Flexibility}: Compared to other existing
     NOMA techniques, such as MUSA~\cite{Yuan2016Multi}, PDMA~\cite{ChenPattern7526461}, SCMA~\cite{Nikopour2013Sparse,Zhiguo2015Mag,Dai2015NOMA}, NOMA
     is conceptually appealing and yields a low-complexity design. In
     fact, the fundamental principles of the aforementioned MA schemes
     and NOMA are very similar, relying on allocating multiple users
     to a single RB. Considering the comparison of NOMA and SCMA as an
     example, SCMA can be regarded as a variation of NOMA which
     integrates appropriate coding, modulation and subcarrier
     allocation.
\end{itemize}


\subsection{Investigating NOMA from an Information Theoretic  Perspective}
Having considered the potential benefits of NOMA,  it is important to investigate its performance gain also from an information theoretic perspective.  In fact, the concept of NOMA may also be interpreted as a special case of SC in the downlink broadcast channel (BC). More particularly,  by using SC, the capacity region of  a realistic imperfect discrete memoryless BC was established by Cover~\cite{cover1972broadcast}. As an extension of~\cite{cover1972broadcast}, Bergmans found the Gaussian BC capacity region of single-antenna scenarios~\cite{Bergmans1974broadcast}. Inspired by~\cite{cover1972broadcast,Bergmans1974broadcast}, several researchers began to explore the potential
performance gain from an information theoretic  perspective \cite{xu2015new,Shieh2016Tcom,Jungho2016Tcom}.
Xu \emph{et al.}~\cite{xu2015new} developed  a new evaluation criterion for quantifying the performance gain of NOMA over OMA. More specifically, considering a simple two-user single-antenna scenario in conjunction with the Gaussian BC, the comparison of TDMA and NOMA in terms of their capacity region was provided in~\cite{xu2015new}. The analytical results showed that NOMA is capable of  outperforming TDMA both in terms of the individual user-rates and the sum rate. In~\cite{Shieh2016Tcom}, Shieh and Huang focused their attentions on examining the capacity region of downlink NOMA
 by systematically designing practical schemes and investigated the gains of NOMA over OMA by designing practical encoders and decoders.
Furthermore, by proposing to use NOMA for relaying broadcast channels (RBC) for the same of achieving a performance enhancement, So and Sung~\cite{Jungho2016Tcom} examined
the achievable capacity region of the RBC upon invoking both decode-and-forward (DF) relaying, as well as compress-and-forward (CF) relaying with/without dirty-paper coding (DPC). Regarding the family of MIMO-NOMA systems, in \cite{liu2016capacity} the achievable capacity region
of multiuser MIMO systems is investigated upon invoking iterative linear minimum mean square error (LMMSE) detection.

\begin{figure}[t!]
    \begin{center}
        \includegraphics[width=3.8in]{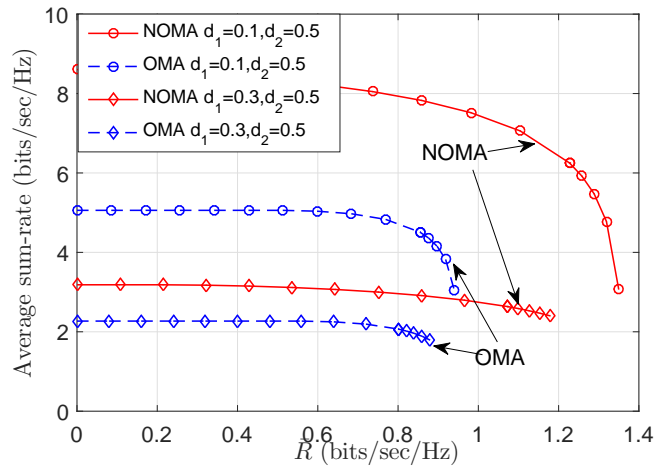}
        \caption{Illustration of the sum-rate versus the minimum rate constraints with full CSIT knowledge for different user distances. The full parameter settings can be found in \cite{Xing2017NOMA}.}
        \label{rate region}
    \end{center}
\end{figure}

In contrast to the above research contributions considering NOMA in additive white Gaussian noise (AWGN) channels, Xing \emph{et al.}  \cite{Xing2017NOMA} investigated the performance of a two-user case of downlink NOMA in fading channels to exploit the time-varying nature on multi-user channels. As shown in Fig. \ref{rate region}, NOMA achieves a superior performance over OMA for different distance settings, where the average sum-rate was maximized subject to a minimum average individual rate constraint.

\begin{figure*}[t!]
    \begin{center}
        \includegraphics[width=16cm]{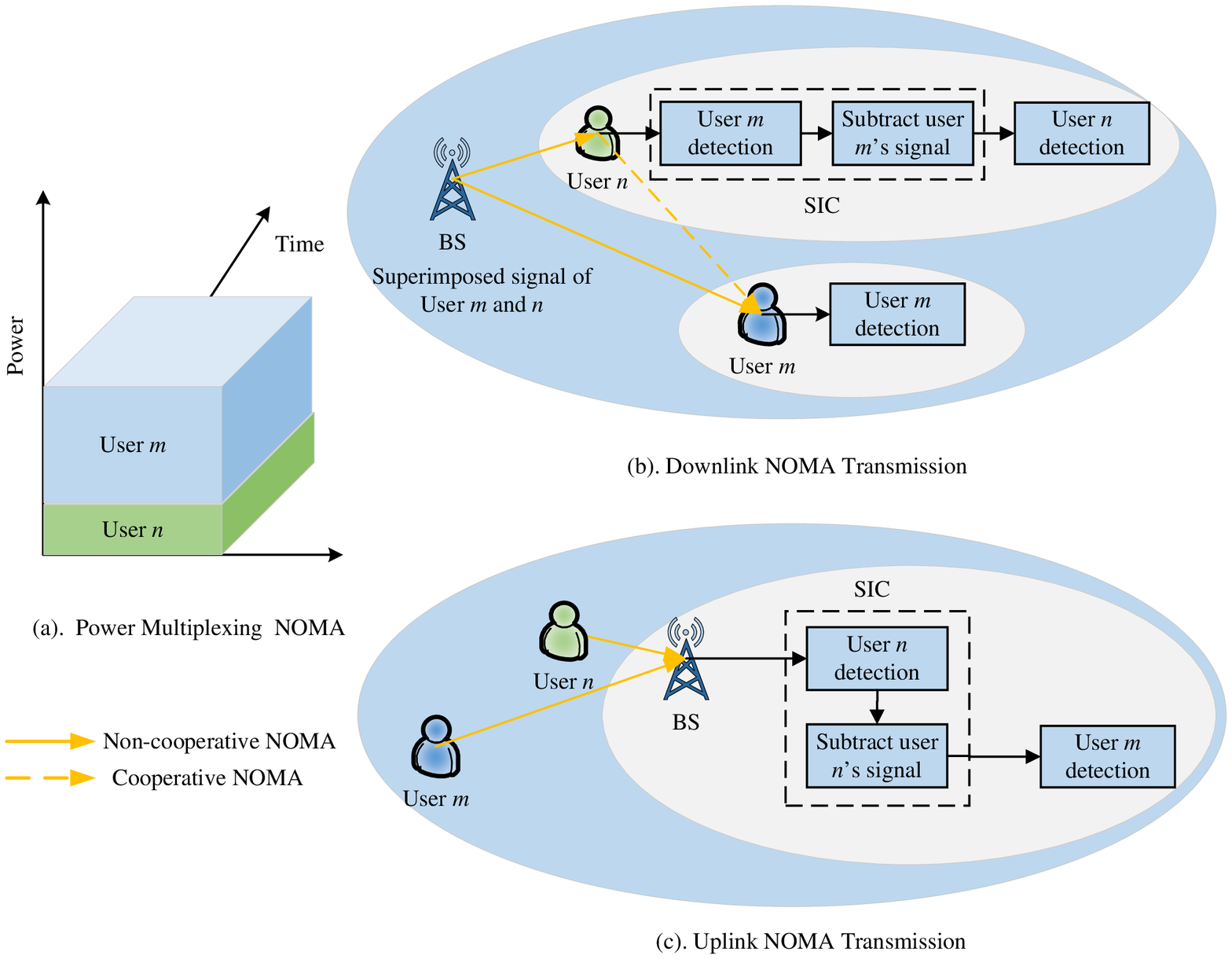}
     \caption{Illustration of NOMA transmission.}
            \label{NOMA basic}
    \end{center}
\end{figure*}
\subsection{Downlink NOMA Transmission}
Downlink NOMA transmission employs the SC technique at the BS for sending a combination of the signals and the SIC technique may be invoked  by the users for interference cancelation, as shown in Fig. \ref{NOMA basic}(b). Numerous valuable contributions have investigated the performance of NOMA in terms of downlink transmission~\cite{Saito:2013,saito2013system,benjebbour2013system,ding2014performance,Timotheou:2015}.
In~\cite{Saito:2013}, a two-user NOMA downlink transmission relying on SIC receivers was proposed. Upon considering a range of further practical conditions in terms of the key link-adaptation functionalities of the LTE,
 the  system-level performance was evaluated in~\cite{saito2013system,benjebbour2013system}. A more general NOMA transmission scheme was proposed in~\cite{ding2014performance}, which considered a BS
communicating with $M$ randomly deployed users. It was demonstrated that NOMA is capable of achieving superior
performance compared to OMA in terms of both its outage probability and its ergodic rate. In~\cite{Timotheou:2015}, the fairness issues were addressed by adopting
appropriate PA coefficients for the $M$ users in a general NOMA downlink transmission scenario.

Motivated by reducing the signalling overhead required for  CSI estimation, some researchers embarked on investigating the performance of downlink NOMA transmissions using partial CSI
at the transmitter~\cite{Yang2016Tcom,shi2015outage,Cui2016NOMA}. More explicitly, Yang \emph{et al.}~\cite{Yang2016Tcom} studied the outage probability of NOMA by assuming either imperfect CSI or
second order statistics based CSI, respectively.  In \cite{shi2015outage}, by assuming the knowledge of statistical CSI, Shi \emph{et al.} investigated the outage performance
of NOMA by jointly considering both the decoding order selection and the PA of the users. In \cite{Cui2016NOMA}, by assuming that only the average CSI was obtained at the BS, Cui \emph{et al.} studied both the optimal decoding order as well as the optimal PA of the users in downlink NOMA systems. Both the transmit power of the BS and  rate-fairness of users were optimized.  By considering only a single-bit feedback of the CSI from each user to the BS, the outage performance of a downlink transmission scenario was studied by Xu \emph{et al.}~\cite{xu2016outage}. Based on the  analytical results derived, the associated dynamic PA optimization problem was solved by minimizing the outage probability. In~\cite{zhang2016energy}, Zhang \emph{et al.} investigated an energy efficiency optimization problem in downlink NOMA systems in conjunction with different data rate requirements of the users. It was shown that NOMA is also capable of outperforming OMA in terms of its energy efficiency.

Apart from  wireless communication systems, the potential use of NOMA in other communication scenarios has also attracted interests. A pair of representative communication scenarios are visible light communication (VLC) \cite{Komine2004visible,Marshoud2016visible,Yin2016NOMAVLC,zhang2016interference} and quantum-aided communication \cite{Hanzo2012Quantum}. To elaborate, in \cite{Marshoud2016visible}, Marshoud \emph{et al.} applied the NOMA technique in the context of VLC downlink networks. By doing so, the achievable throughput was enhanced. It is worth pointing out that although the potential performance enhancement is brought by invoking NOMA into VLC networks, the hardware implementation complexity is also increased at transmitters and receivers as SC and SIC techniques are adopted. In \cite{Botsinis2016Quantum}, Botsinis \emph{et al.} considered quantum-assisted multi-user downlink transmissions in NOMA systems, where a pair of bio-inspired algorithms were proposed.

\subsection{Uplink NOMA Transmission}
In uplink NOMA transmission, multiple users transmit their own uplink signals to the BS in the same  RB, as shown in Fig. \ref{NOMA basic}(c). The BS detects all the messages of the users with the aid of SIC.  Note that there are several key differences between uplink NOMA and downlink NOMA, which are listed as follows:

\begin{itemize}
\item \textbf{Transmit Power:} In contract to downlink NOMA, the transmit power of the users in uplink NOMA does not have to be different, it depends on the channel conditions of each user. If the users' channel conditions are significantly different, their received SINR can be rather different at the BS, regardless of their transmit power.
  \item \textbf{SIC Operations:}  The SIC operations and interference experienced  by the users in the uplink NOMA and downlink NOMA are also rather different. More specifically, as shown in Fig. \ref{NOMA basic}(b) for downlink NOMA, the signal of User $n$ is decontaminated from the interference imposed by User $m$, which is achieved by first detecting the stronger signal of User $m$, remodulating it and then subtracting it from the composite signal. It means that SIC operation is carried out on strong user in downlink for canceling the weak user's interference.  By contrast,  in uplink NOMA, SIC is carried out at the BS to   detect strong User $n$  first by treating User $m$ as interference, as shown in Fig. \ref{NOMA basic}(c). Then it remodulates the recovered signal and subtracts the interference imposed by User $n$ to detect User $m$.
  \item \textbf{Performance Gain:} The performance gain of NOMA over OMA is different for downlink and uplink.  Fig. 1 of \cite{shin2016non} illustrates the capacity region of NOMA and OMA both for downlink and uplink. The capacity region of NOMA is outside  OMA, which means that the use of NOMA in downlink has superior performance in terms of throughput. While in uplink, NOMA mainly has the advantages in terms of fairness, especially compared to that that OMA with power control.
\end{itemize}

Recently, there have been less research contributions on uplink NOMA than on downlink NOMA~\cite{higuchi2015non,al2014uplink,alreceiver2016uplink,Chen2015uplink,Zhang2016uplink,Zhiguo_general:2015,WPT_NOMA_2015}.
The downlink and uplink NOMA  throughput regions were quantified by Higuchi and Benjebbour~\cite{higuchi2015non} for both symmetric and
asymmetric channels. They demonstrated that the performance benefits of  NOMA over OMA become more significant, when the channel conditions of the users are
more different.
Al-Imari \emph{et al.}~\cite{al2014uplink} designed a novel NOMA scheme for the classic OFDMA  uplink. The efficiency gains of the proposed scheme over the conventional OMA scheme were quantified both in terms of fairness and spectral efficiency.
As an extension of~\cite{al2014uplink}, a novel iterative multiuser detection was designed in~\cite{alreceiver2016uplink} for further enhancing the
performance of the NOMA uplink. In an effort to reduce the implementation complexity, Chen \emph{et al.}~\cite{Chen2015uplink} conceived a user-pairing policy for the  NOMA uplink.
 In~\cite{Zhang2016uplink}, Zhang \emph{et al.} proposed a novel power control strategy for the NOMA uplink
and investigated both the outage probability and the delay-limited sum-rate.  In~\cite{WPT_NOMA_2015}, a wirelessly powered uplink NOMA transmission scheme was
investigated by Diamantoulakis \emph{et al.} by applying a harvest-then-transmit style EH protocol.
A general MIMO-NOMA framework applying stochastic geometry both for downlink and uplink transmission was proposed by Ding \emph{et al.} \cite{Zhiguo_general:2015}.

\subsection{Discussions and Outlook}
Given the increasing number of research contributions on NOMA, its advantages are becoming increasingly clear, especially in terms of its bandwidth efficiency, energy efficiency and fairness. Stochastic geometry constitutes a powerful mathematical and statistical tool, which is capable of capturing the topological randomness of the networks and hence
provides tractable analytical results for the average network behavior \cite{Stoyan}. This is particularly essential in large-scale networks supporting a large number of randomly distributed BSs and mobile users. At the time of writing the
stochastic geometry based modelling of  NOMA networks is still in its infancy \cite{ding2014performance,yuanwei_JSAC_2015,Zhiguo_general:2015,Qin2016PLS_NOMA,Liu2016TVT,Zhiguo2016mmWave,tabassum2016modeling}. The derivation of closed-form expressions for obtaining deep insights
remains an open challenge. Hence further research efforts are required for analyzing the average performance of multi-user multi-cell NOMA networks \cite{Shin2017NOMA,han2014energy} in order to provide inspiration both for the long-term investigations in academia and practical guidelines for industry.

Another important issue to be addressed in the NOMA context is that of the
ordering of users as required by SIC. Order statistics \cite{order} can be a helpful tool to assist the associated performance analysis for the sake of obtaining insightful guidelines.

Most of the existing contributions are mainly focused on the theoretical performance analysis or optimization. Most of the implementation issues, such as the effect of imperfect CSI on both SISO-NOMA \cite{Jingcui2016SPL} and on MISO-NOMA \cite{Zhang2016TVT,Cumanan2017COML,Jiayin2016SPL}, on
limited channel feedback \cite{liu2017downlink,Xu2016back}, on efficient receiver designs \cite{tomasi2017low}, and its combination with advanced adaptive coding and modulation schemes, etc are still open areas.
Correspondingly, further research is expected in the aforementioned areas to conceive a holistic architecture for NOMA.
\section{NOMA Combined with Multiple Antennas Techniques}\label{section:MIMO-NOMA}
Multiple antenna techniques are of significant importance, since they offer the extra dimension of the spatial domain, for further performance improvements.   The application of multiple antenna techniques in NOMA has attached substantial interest both from academia~\cite{higuchi2015non,Sunqi2015,ding2015mimo,Jinho:2015,Zhiguo_general:2015,Fainan2015TSP,Jinho2016MIMO,Zhiguo2016IoT,kim2013non,Soma2016Uplink,Liu2016MIMO+NOMA,Chen2016TSP,Chen2016NOMA,yu2016antenna,Ali2017MIMONOMA} and from industry~\cite{higuchi2013non,higuchi2015non,benjebbour2013concept,Saito:2013}. The distinct NOMA features such as channel ordering and PA inevitably require special attention in the context of multiple antennas. More specifically, in contrast to the SISO-NOMA scenarios whose channels
are all scalars, the channels of MIMO-NOMA scenarios are represented in form of matrices, which makes the power-based ordering of users rather challenging. As a consequence,
conceiving an appropriate beamforming/precoding design is essential for multi-antenna aided NOMA systems. NOMA relying on beamforming (BF) constitutes an efficient technique of improving the bandwidth efficiency by exploiting both the power domain and the angular domain. There are two popular MIMO-NOMA designs, namely the 1) Cluster-based (CB)
MIMO-NOMA design; and the 2) Beamformer-based (BB) MIMO-NOMA design, which will be introduced in the following.
Table \ref{table:MIMO NOMA} summarizes some of the existing contributions on NOMA applying multiple antennas and illustrates their comparison.

\begin{table*}[htbp]
\caption{Important contributions on multi-antenna aided NOMA. ``DL" and ``UL" represent downlink and uplink, respectively. ``BF" and ``OP" represent beamforming and outage probability, respectively.``SU" and ``MU" represent two-user and multi-user cases. The ``sum-rate gain" implies that the gain brought by invoking NOMA technique over conventional OMA technique in terms of sum-rate.
}
\begin{center}
\centering
\begin{tabular}{|l|l|l|l|l|}
\hline
\centering
 \textbf{Ref.} &\textbf{Scenarios} & \textbf{Direction} & \textbf{Main Objectives} & \textbf{Techniques/Characteristics} \\
\hline
\centering
 \cite{Jinho:2015} & MU-MISO & DL &  Transmit power & Two stage BF \\
\hline
\centering
 \cite{higuchi2015non} & MU-MISO & DL/UL & Throughput & Less constraint at the BS \\
\hline
\centering
 \cite{Fainan2015TSP} & MU-MISO & DL  & Sum-rate & Ordered vector constraints \\
\hline
\centering
\cite{higuchi2013non} &  MU-MIMO & DL  & Sum-rate & Random BF\\
\hline
\centering
\cite{ding2015mimo} &  MU-MIMO & DL & Sum-rate gain & No need for perfect CSIT at the BS\\
\hline
\centering
  \cite{Zhiguo_general:2015}& MU-MIMO & DL/UL  & OP  & Obtaining larger diversity gain \\
\hline
\centering
  \cite{Zhiguo2016IoT} & SU-MIMO & DL & OP & Creating channel differences artificially\\
\hline
\centering
\cite{Sunqi2015} & SU-MIMO & DL & Ergodic capacity & Optimal/sub-optimal power allocation\\
\hline
\centering
  \cite{Jinho2016MIMO} & SU-MIMO & DL & Sum-rate & Low complexity decoding\\
\hline
\centering
  \cite{Ding2016SPL} & MU-MIMO & DL & OP & Limited feedback requirements\\
\hline
\centering
  \cite{Yuanwei2016NOMA} & MU-MIMO & DL& Max-min rate & Addressed max-min fairness issue \\
\hline
\centering
  \cite{Yuanwei2017TWC} & MU-MISO & DL& Secrecy OP & Generating artificial noise\\
\hline

\end{tabular}
\end{center}
\label{table:MIMO NOMA}
\end{table*}
\subsection{Cluster-Based MIMO-NOMA}
\begin{figure}[t!]
    \begin{center}
        \includegraphics[width=8cm]{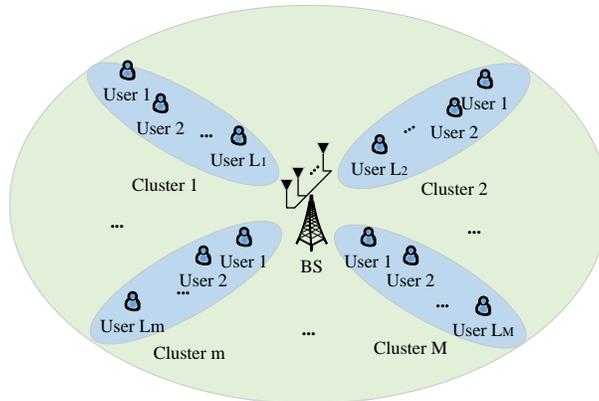}
     \caption{Illustration of cluster-based MIMO-NOMA.}
            \label{NOMA BF}
    \end{center}
\end{figure}

One of the popular NOMA designs is associated with the cluster-based structure, partitioning users into several different clusters. Explicitly, as shown in Fig. \ref{NOMA BF}, the NOMA users are partitioned into $M$ clusters and each cluster consists of $L_m$ users, where $m \in \left\{ {1,2,...,M} \right\}$.
Then we design appropriate beams for the corresponding clusters. Upon applying effective transmit precoding and detector designs, it becomes possible to guarantee that the beam associated with a particular cluster is orthogonal to the channels of users in other clusters. Hence the inter-cluster interference can be efficiently suppressed. When considering each cluster in isolation, there is a  difference among the users' channel conditions, hence we are faced again with the conventional NOMA scenarios. Thus, SIC can be readily invoked for mitigating the intra-cluster interference between users of the same cluster. Recently, many important research contributions investigated beamforming aided NOMA \cite{Jinho:2015,higuchi2015non,ding2015mimo,Zhiguo_general:2015,Zhiguo2016IoT,Yuanwei2016NOMA}.


\begin{figure}[t!]
    \begin{center}
        \includegraphics[width=9cm]{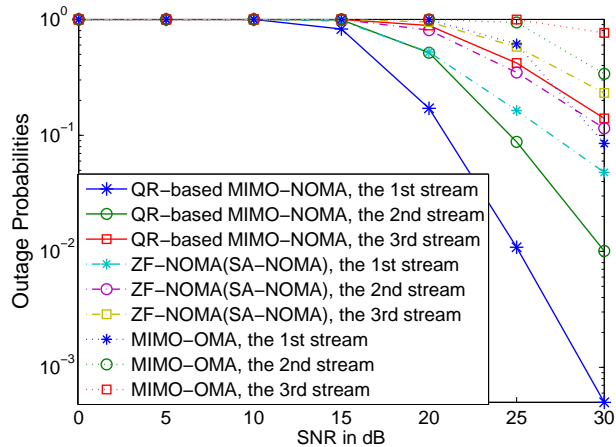}
     \caption{Comparison of MIMO systems with various multiple access techniques using a two-user case.  $M=N=3$, where $M$ is the number of antennas at transmitters, $N$ is the number of antennas at receivers. As such, each user is to receive three data streams from the BS. $R_m = 1.2$ bit per channel use (BPCU) and $R_n=5$ BPCU are the targeted data rates of User $m$ and User $n$, respectively. }
            \label{MIMO NOMA comparision}
    \end{center}
\end{figure}

Specifically, Choi~\cite{Jinho:2015} proposed two-stage multicast zero-forcing (ZF)-based beamforming for downlink inter-group/cluster interference mitigation, where the total transmitt power of each group/cluster was minimized during the second stage. Higuchi and Benjebbour~\cite{higuchi2015non} invoked receive beamforming at the NOMA users and a transmit beamformer at the BS. Higuchi and  Kishiyama~\cite{higuchi2013non} then proposed a novel scheme, which combined open-loop  random beamforming in conjunction with intra-beam SIC for downlink NOMA transmission. However, random beamforming fails to guarantee a constant QoS at the users' side. To overcome this limitation, in \cite{ding2015mimo}, Ding \emph{et al.} proposed a TPC and detection scheme combination for a cluster-based  downlink MIMO-NOMA scenario relying on fixed PA. By adopting this design, their MIMO-NOMA system can be decomposed into several independent single-input single-output (SISO) NOMA arrangements. Furthermore, in order to establish a more general framework  considering both downlink and uplink MIMO-NOMA scenarios, the so-called signal alignment (SA) technique was proposed in~\cite{Zhiguo_general:2015}. Stochastic geometry based tools were invoked to model the impact of the NOMA users' locations~\cite{Zhiguo_general:2015}. In contract to the research contributions in \cite{ding2015mimo,Zhiguo_general:2015}, which are inter-cluster interference free design, an inter-cluster interference allowance design for CB MIMO-NOMA was proposed in \cite{Ali2017access}. More particularly, a user who experiences the highest channel condition was selected as a cluster-head and was capable of completely canceling all  intra-cluster interference in \cite{Ali2017access}.
Note that the existing NOMA designs have routinely relied on assuming different channel conditions for the different users, which is however a somewhat restrictive assumption. In order to circumvent this restriction, Ding \emph{et al.}~\cite{Zhiguo2016IoT} designed a new MIMO-NOMA scheme, which distinguishes the users according their QoS requirements with particular attention on IoT scenarios for the sake of supporting the SIC operation. Furthermore, they compared this new  MIMO-NOMA design to two NOMA schemes, which order users according to the prevalent channel conditions. More particularly, the ZF-NOMA scheme of \cite{ding2015mimo} and the SA-NOMA scheme~\cite{Zhiguo_general:2015} were used as benchmarks in~\cite{Zhiguo2016IoT}. Fig. \ref{MIMO NOMA comparision} illustrates the outage probability defined as the probability of erroneously detecting the message intended for User $m$ in the $i$-th data stream, $i=1,2,3$ at User $n$, where the QR decomposion is used to augmenting the differences between the users' effective channel conditions according to the associated QoS requirements.  As shown in Fig. \ref{MIMO NOMA comparision},  the QR-based MIMO-NOMA scheme is capable of outperforming both ZF-NOMA and SA-NOMA\footnote{Note that when $M=N$, ZF-NOMA achieves the same performance as SA-NOMA~\cite{Zhiguo2016IoT}.} as well as MIMO-OMA\footnote{Fig. \ref{MIMO NOMA comparision} is focused on the performance of User $n$, since the QoS requirements have been guaranteed with the aid of appropriate PA~\cite{Zhiguo2016IoT}.}, since it exploits the heterogeneous QoS requirements of different users and applications. In \cite{Yuanwei2016NOMA}, the fairness issues of the MIMO-NOMA scenario were addressed by applying appropriate user allocation algorithms among the clusters and dynamic PA algorithms within each cluster.

\subsection{Beamformer-Based MIMO-NOMA}
\begin{figure}[t!]
    \begin{center}
        \includegraphics[width=9cm]{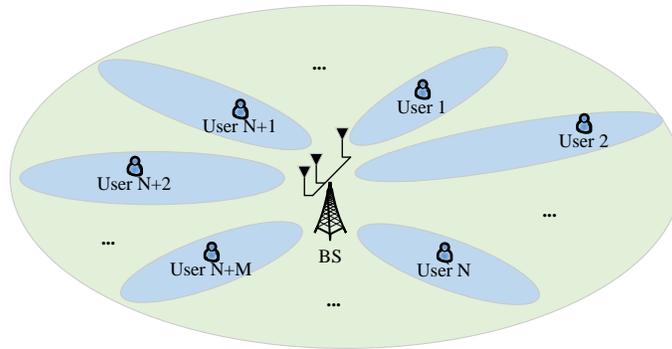}
     \caption{Illustration of beamformer-based MIMO-NOMA.}
            \label{NOMA BB}
    \end{center}
\end{figure}

Another technique of implementing MIMO-NOMA is to assign different beams to different users, as shown in Fig. \ref{NOMA BB}. By doing so, the QoS can be satisfied
by calculating the beamformer-weights in a predefined order, commencing with the most demanding QoS requirement. By adopting this approach, several contributions have been made in terms of MIMO-NOMA. Considering the illustration of Fig. \ref{NOMA BB} as an example, User $1$ to User $N$ occupy the same RB, similarly to user $(N+1)$ to user $(N+M)$. Again, within the same RB we may employ SIC at each user, according to the particular ordering of the different users' received signal power.  In \cite{Fainan2015TSP}, Hanif \emph{et al.} solved the downlink sum-rate maximization problems, which resulted in obtaining the corresponding optimal TPC vectors.
Sun \emph{et al.}~\cite{Sunqi2015} first investigated
the power optimization problem constructed for maximizing the ergodic capacity and then showed that their proposed MIMO-NOMA schemes are capable of achieving significantly better performance than OMA.
In an effort to reduce the decoding complexity imposed at the users, a layered transmission based MIMO-NOMA scheme was proposed by Choi~\cite{Jinho2016MIMO}, who also  investigated the associated PA problem. It was demonstrated that upon invoking this layered transmission scheme, the achievable sum rate increases linearly with the number of antennas.

\subsection{Massive-MIMO-NOMA}
Massive MIMO may be considered as one of the key technologies \cite{andrews2014will} in 5G systems as a benefit of improving both the received SNR and the bandwidth efficiency. It was shown in~\cite{Larsson2014CM} that massive MIMO is capable of substantially increasing both the capacity as well as the energy efficiency. These compelling benefits of massive MIMO sparked off the interest of researchers also in the context of NOMA. In \cite{Ding2016SPL}, Ding and Poor conceived a two-stage TPC design for implementing massive-MIMO-NOMA. More particularly, a beamformer was
adopted for serving to a cluster of angularly similar users and then they decomposed the MIMO-NOMA channels into a number of SISO-NOMA channels within the same cluster. A one-bit CSI feedback scheme was proposed for maintaining a low feedback overhead and a low implementation  complexity.

\subsection{Discussions and Outlook}
Although numerous research contributions have been made in the field of MIMO-NOMA scenarios, most of the existing treatises assumed that the perfect CSI is known at the transmitter. Under this assumption, the TPC matrix may be readily designed according to the ordered channels between the users and the BS based on their received power. However, in practical wireless communication systems, obtaining the channel state information at the transmitter (CSIT) is not a trivial problem, which requires the classic pilot based training process. This is particularly important for massive-MIMO-NOMA scenarios, where a large number of antennas is used at the BS, hence potentially imposing an excessive pilot-overhead and complexity. Additionally, when the Doppler-frequency is doubled, so is the pilot-overhead and the computational complexity. Although a low-complexity massive-MIMO-NOMA scheme relying on an one-bit feedback flag was proposed in \cite{Ding2016SPL}, naturally, the performance was degraded compared to the case of perfect CSIT. Motivated by all the aforementioned problems, efficient CSIT estimation techniques are required for supporting massive-MIMO-NOMA systems. By exploiting the sparsity of the user's channel matrix, compressive sensing (CS) can be invoked for reducing the pilot-overhead \cite{Rao2014TSP}, as well as for reducing the feedback overhead transmitted from the downlink receivers to the BS. In the CS based CSI estimation conceived for massive-MIMO-NOMA,  machine learning may be invoked both for identifying the measurement matrix design of CS and for the sparse domain selection process of channel estimation in massive-MIMO-NOMA. Explicitly, this is another valuable research topic.

The channel estimation problem can also be jointly considered with the data detection, where soft extrinsic-information is exchanged between the receiver blocks, such as the channel estimator, data detector and channel decoder. In this context it is worth emphasizing that once some of the soft estimates become sufficiently reliable, then they can be subjected to hand-decisions and then used as additional pilots in the context of decision directed channel estimation \cite{hanzo2007ofdm}. Finally, non-coherent MIMO-NOMA dispensing with channel estimation has to be investigated. Additionally, when considering the practical implementation issues, the co-existence of NOMA and MIMO inevitably necessitates the redesign of receivers because of the distinct characteristics of SIC.
\section{Interplay between NOMA and Cooperative Communications}


Due to the hostile characteristics of wireless channels, the attenuation of the signals may vary  dramatically during their transmission. As a counter-measure, cooperative communication is a particularly attractive technique of
offering cooperative diversity gains, hence extending the network's coverage~\cite{laneman2004cooperative,Nosratinia2004Cooperative}. In this section, we summarize the NOMA contributions on cooperative communications, mainly from the perspective of promising cooperative NOMA techniques and the application of NOMA in cooperative
networks.
\subsection{Cooperative NOMA}
The key idea behind cooperative NOMA is to rely on strong NOMA users acting as DF relays to assist weak NOMA users. Still considering
the two-user downlink transmission of Fig. \ref{NOMA basic}(b) as our example, cooperative NOMA requires two time slots for its transmission. The
first slot is for the direct transmission phase, which is the same as the non-cooperative NOMA of Fig. \ref{NOMA basic}(b) indicated by solid lines.
During the second time slot, which is the cooperative phase, User $n$ will forward the decoded message to User $m$ by invoking the DF relaying protocol of Fig. \ref{NOMA basic}(b) indicated by the dashed line.
Proposed by Ding \emph{et al.}~\cite{Ding2015cooperative}, this novel concept intrigued researchers, since cooperative NOMA fully benefits from SIC and DF decoding.
In~\cite{Ding2015cooperative}, a general downlink NOMA transmission scenario was considered, where the BS supported $M$ users with the aid of cooperative
NOMA protocols relying on $M$ slots. In an effort to seek a more efficient cooperative NOMA protocol, Liu \emph{et al.}~\cite{yuanwei_JSAC_2015}
proposed a new EH assisted cooperative NOMA scheme. A sophisticated stochastic geometry based model was invoked for evaluating the system's performance
and user pairing was adopted for reducing the implementation complexity. Compared to conventional NOMA, the key advantages of cooperative NOMA transmissions can be summarized as follows:

\begin{itemize}
  \item \textbf{Low System Redundance:} Again, upon applying SIC techniques in NOMA, the message of the weak user has already been decoded at the
  strong user, hence it is natural to consider the employment of the DF protocol for weak signal. Explicitly, the weak signal can be remodulated
  and retransmitted from a  position closer to the destination.
  \item \textbf{Better fairness:} A beneficial feature of cooperative NOMA is that the reliability of the weak user is significantly improved. As a consequence,
  the fairness of NOMA transmission can be improved~\cite{Timotheou:2015}, particularly in the scenarios when the weak user is at the edge of the cell illustrated by the BS.
  \item \textbf{Higher diversity gain:} Cooperative NOMA is capable of achieving an improved diversity gain for the weak NOMA user, which is an effective technique of overcoming multi-path fading. It was analytically demonstrated~\cite{Ding2015cooperative} that the diversity gains of the weak users in cooperative NOMA are the same as those of the conventional cooperative networks,
  even for using EH relays~\cite{yuanwei_JSAC_2015}.
\end{itemize}

\begin{figure}[t!]
    \begin{center}
        \includegraphics[width=3.5in]{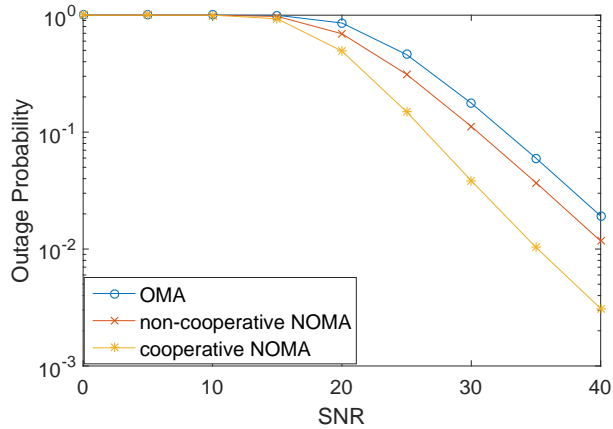}
        \caption{Performance comparison between cooperative NOMA and non-cooperative NOMA. The BS is located at (0, 0), User $m$ is located at (4, 0), User $n$ is located at (1, 0),  the path loss exponent is 3,  the power allocation coefficient for User $m$ and User $n$  are (0.8, 0.2).  The targeted data rate is 1 bits per channel use (BPCU).}
        \label{cooperative NOMA vs non}
    \end{center}
\end{figure}
Fig. \ref{cooperative NOMA vs non} illustrates the superior performance of cooperative NOMA over non-cooperative NOMA as well as over OMA in terms of its outage probability. It is noted that cooperative NOMA achieves a higher diversity gain than non-cooperative NOMA and OMA, which demonstrates the effectiveness of cooperative NOMA, as aforementioned. Note that cooperative NOMA constitutes one of many techniques of improving the transmission reliability of NOMA networks, especially for the weak NOMA users who have poor channel conditions. There are also other techniques for enhancing the performance of NOMA networks, which will be detailed in the following subsection. It is also worth pointing out that the performance of cooperative NOMA is related to the error propagation issue for SIC, which will be introduced in Section VII.

\subsection{NOMA in Cooperative Transmission Based Networks}
Cooperative communication and NOMA techniques mutually support each others,
hence the joint action of both techniques further improves the  cooperative network performance~\cite{Men2015,Men2015CL,Jinho2014Comp,Tian2016CoMP,Kim2015CoMP,Xu2015GC,Vien2015CRAN}.
Relaying and CoMP transmission are capable of improving cooperative networks.

\subsubsection{Relay-aided NOMA Transmission}
Relaying has recently attracted considerable attention as a benefit of its network coverage extension~\cite{laneman2004cooperative}.
In this context, Men and Ge investigated the outage performance of the single-antenna AF relay aided NOMA downlink \cite{Men2015}. By contrast,  in \cite{Men2015CL} multiple  antennas were used. The potential gains of NOMA over OMA were quantified in both scenarios. As a further development, a DF based two-stage relay selection protocol was proposed in \cite{ding2016relay}, which relied on maximizing the diversity gain and minimizing the outage probability. In \cite{duan2016use}, Duan \emph{et al.} proposed a novel two-stage PA scheme for a dual-hop relay-aided NOMA  system, where the destination jointly decoded the information received both from the source and from the relay by applying the classic MRC technique. In~\cite{Kim2015CoMP}, Kim and Lee considered a coordinated direct and relay-aided transmission scheme, where the BS simultaneously transmitted both to a nearby user and to a relay by invoking NOMA techniques
during the first phase, while reaching a distant user with the aid of the relay.

\subsubsection{Multi-Cell NOMA with Coordinated Multipoint Transmission}

When considering multi-cell scenarios, the performance of the  cell-edge users is of particular concern. This is particularly important for downlink NOMA, since the SIC operations are usually carried out for cell-center users rather than for cell-edge users. Hence the cell-edge users may not be well served. CoMP transmissions constitute an effective technique of improving the performance of cell-edge users. The key concept of CoMP in multi-cell NOMA is to enable multiple BSs to carry out coordinated beamforming or joint signal processing for the cell-edge users \cite{ali2017coordinated}. There are several research contributions in the context of handling the inter-cell interference in NOMA networks \cite{han2014energy,Jinho2014Comp,Shin2017NOMA,Tian2016CoMP}, as detailed below.


In \cite{han2014energy}, Han \emph{et al.} extended a single cell NOMA to multi-cell network-NOMA. A new precoding approach was proposed for mitigating the inter-cell interference, which lead to an enhanced spectral efficiency and energy efficiency. In~\cite{Jinho2014Comp}, Choi incorporated NOMA into CoMP for the sake of attaining a bandwidth efficiency improvement.
A new coordinated superposition coding scheme relying on Alamouti's space-time code was proposed. As a further advance, Shin \emph{et al.} \cite{Shin2017NOMA} investigated the performance of multi-cell MIMO-NOMA networks, applying coordinated beamforming for dealing with the inter-cell interference in order to enhance the cell-edge users' throughput.  Tian \emph{et al.}~\cite{Tian2016CoMP} conceived an opportunistic NOMA scheme for CoMP systems and compared it to the conventional joint-transmission based
NOMA.
In \cite{Vien2015CRAN}, Vien \emph{et al.} proposed a NOMA based PA policy for downlink cloud radio access network (C-RAN) scenarios, which can also be regarded as a coordinated
transmission scenario, where the BSs jointly form a cloud. In \cite{Francisco2017NOMA}, Martin \emph{et al.} proposed a novel NOMA-aided C-RAN network, associated with using stochastic geometry. It revealed that the proposed framework is capable of substantially enhancing the performance of cell-edge users.

\begin{table*}[htbp]
\caption{Contributions on the interplay of NOMA and cooperative communications}
\begin{center}
\centering
\begin{tabular}{|l|l|l|l|l|l|}
\hline
\centering
 \textbf{Ref.} &\textbf{Model Topology} & \textbf{Direction} & \textbf{Techniques}& \textbf{Main Metrics} & \textbf{Characteristics} \\
\hline
\centering
 \cite{Ding2015cooperative} & Idealized topology & DL & DF & OP+ergodic rate & Maximum diversity gain achievable \\
\hline
\centering
 \cite{yuanwei_JSAC_2015} & Stochastic geometry & DL & DF & OP & Wireless powered spatial randomly relays \\
\hline
\centering
 \cite{Men2015CL} & Idealized topology & DL & AF & OP+ergodic rate & Multiple-antenna aided relays \\
\hline
\centering
\cite{ding2016relay} &  Idealized topology & DL & DF & OP & Novel two-stage relay selection policy\\
\hline
\centering
 \cite{han2014energy}& Idealized topology & DL  & CoMP & SE + EE  & Zero-forcing like distributed precoding\\
\hline
\centering
  \cite{Jinho2014Comp}& Idealized topology & DL  & CoMP & Sum-rate  & Using Alamouti code at coordinated BSs\\
\hline
\centering
  \cite{Shin2017NOMA}& Idealized topology & DL  & CoMP & Sum-rate  & MIMO-NOMA coordinated beamforming\\
\hline
\centering
  \cite{Tian2016CoMP} & Idealized topology & DL& CoMP & OP+sum-rate & Joint transmission/Opportunistic NOMA in CoMP\\
\hline
\centering
\cite{Kim2015CoMP} & Idealized topology & DL& CoMP & OP+ergodic rate & Each stream achieves a diversity order of one\\
\hline
\centering
  \cite{Vien2015CRAN} & Idealized topology& DL& CRAN & Throughput & All the BSs are controlled by a cloud\\
\hline
\centering
  \cite{duan2016use} & Idealized topology & DL& DF & Sum-rate & Novel two-stage power allocation\\
\hline
\centering
  \cite{Xinwei2017FD} & Idealized topology & DL& DF & OP & Invoking full-duplex technique at relays\\
\hline
\centering
\cite{Francisco2017NOMA} & Stochastic geometry & DL& CRAN & OP & Cluster point process based model\\
\hline
\end{tabular}
\end{center}
\label{table:cooperative NOMA}
\end{table*}

\subsection{Discussions and Outlook}
Table \ref{table:cooperative NOMA} summarizes all the above-mentioned research contributions relying on the interplay between NOMA and cooperative communications. In Table \ref{table:cooperative NOMA}, ``DL" and ``OP" represent the downlink and outage probability, respectively. The terminology of ``Idealized topology" in the table indicates that the users may not be randomly positioned and that the path loss is assumed to be constant. The aforementioned literature has investigated the potential performance enhancements of NOMA with the aid of cooperative techniques. Nonetheless, there are several open research opportunities. For example, although employing relays is capable of improving the reception reliability, whilst extending the coverage area, it requires an extra time slot for relaying. Developing full-duplex relays for NOMA networks constitutes a promising research direction for eliminating the requirement of an extra slot \cite{ZhangTVT2016,Zhong2016FDNOMA,Xinwei2017FD}. However, invoking full-duplex relays will require the elimination of both the self interference and of the inter-user interference.

Additionally, the existing research contributions on C-RAN NOMA are still in their infancy.  It is worth pointing out that C-RAN techniques are capable of efficient interference management and large scale data control. This is particularly important for large-scale C-RAN NOMA networks, where sophisticated interference management (e.g. distributed beamforming design), dynamic user association and efficient PA can be jointly
considered.
\section{Resource Management in NOMA Networks}
One of the main challenges in NOMA networks is to strike an attractive compromise between the bandwidth efficiency and energy efficiency of the networks by intelligently controlling
the PA of the superimposed signals \cite{song2016resource}, dynamically scheduling the users for the sub-channels or by forming spatially correlated clusters.
Motivated by this, in this section, we provide a comprehensive review of the existing resource allocation contributions, mainly with a special emphasis on power control and
user/resource allocation.
\subsection{Power Control for NOMA}
Power control/allocation has been a pivotal research direction throughout each generation of networks, since inappropriately allocating power
 among users will inevitably increase the overall energy consumption whilst inflicting extra interference and hence degrading the overall performance of the networks. Again, in NOMA more power is required for users with poor channel conditions and less power for users with better channel conditions, which guarantees fairness among NOMA users. 

Numerous valuable contributions have been published on the PA problem in NOMA, which may be divided into two main categories, optimal PA from the perspective of all users and
cognitive PA from the perspective of the weak users, which will be detailed in the following two subsections.

\subsubsection{Optimal power allocation}
The contributions on NOMA which investigate optimal PA can be also classified into two main categories: i) \emph{Single-channel/carrier PA}; and ii) \emph{multiple-channel/carrier/cluster PA}. In this subsection, we mainly focus our attention on single-channel/carrier PA, while multiple-channel/carrier/cluster power
allocation will be introduced in Section~\ref{multiple channel power allocation}.

The aim of PA in single channel/carrier scenarios is to optimize the individual/sum rate
whilst giving cognizance to the  the \emph{fairness issues}. Compared to OMA, optimal PA in NOMA imposes more constraints associated with channel-quality and power-based ordering
in an effort to guarantee fairness by allocating reasonable data rates to weak users as well. The recent research contributions on the fairness issues of NOMA can be summarized as follows:

\begin{itemize}
 \item \textbf{Ordered power allocation}: An simple yet effective approach to guarantee the fairness of NOMA systems is to allocate more power to users with poor channel conditions.  By doing so, weak users
  can still achieve adequate rates. It is worth pointing out that adding these transmit power based ordering constraints will impose additional complexity on
   the  optimization problem formulated, especially for multi-antenna aided NOMA scenarios \cite{Jinho:2015,Fainan2015TSP}.
  \item \textbf{Max-min rate-fairness}: The max-min PA problem maximizes the rate of the weakest of all NOMA users~\cite{Timotheou:2015,Yuanwei2016NOMA,Cui2016NOMA}.
  In \cite{Timotheou:2015}, the PA strategy was investigated under the scenarios of knowing either the instantaneous CSI or the average CSI.  Note that by adopting
      the max-min rate as our objective function we can guarantee a certain grade of rate-fairness, but at the price of sacrificing the system's sum-rate.
   \item \textbf{Proportional fairness}: Proportional fairness (PF) is capable of maximizing the geometric mean of user rates. In NOMA scenarios, a feasible proportional fairness policy is to schedule users based on the instantaneous user rates, whilst additionally guaranteeing a certain long-term-average target rate~\cite{Liu2016ICC,Mei2016ICC}.
    \item \textbf{$\alpha$ utility function}: The $\alpha$ utility function constitutes  a generalization of proportional fairness and max-min fairness. The definition of the $\alpha$ utility function is as follows~\cite{Mo2000fariness}:
 \begin{align}\label{alpha utility}
{f_\alpha }\left( x \right) = \left\{ \begin{array}{l}
 \log x,{\kern 1pt} \;\;\;\;\;\;\;\;\;\;\;\;\;\;\;\;if\;\alpha  = 1\; \\
 {x^{1 - \alpha }}/\left( {1 - \alpha } \right),\;\;otherwise. \\
 \end{array} \right.
\end{align}
Note that for $\alpha=1$, Eq \eqref{alpha utility} reduces to that of proportional fairness, while for $\alpha  \to \infty $,  it results in the max-min fairness optimization problem. Hence, the $\alpha$ utility function is capable of treating the fairness issues of NOMA in a generic way.
   \item \textbf{Weighted sum rate}: The key idea behind the weighted sum rate is to consider an additional positive weighing factor for each user's achievable rate, which reflects the priority of each user in the context of resource allocation~\cite{sun2016optimal}. By doing so, a certain grade of fairness can be achieved from the specific perspective of the media access control (MAC) layer.
  \item \textbf{Jain's fairness index comparison}: Jain's fairness index \cite{jain1984quantitative}, is widely used for striking a tradeoff between the
   sum-rate and fairness of a communication system. Motivated by this, many NOMA contributions characterized their PA in terms
  of  Jain's fairness index \cite{al2014uplink,lei2016power,WPT_NOMA_2015}. However, Jain's fairness index
  is only a metric of evaluating a system's fairness, which cannot prevent the weak users from having low rates.
  More intelligent algorithms are required for fair PA  among  the users.
\end{itemize}

\begin{table*}[htbp]\tiny 
\caption{Strategies considering the fairness issues of NOMA}
\begin{center}
\centering
\begin{tabular}{|l|l|l|l|l|l|}
\hline
\centering
 \textbf{Fairness Strategy} &\textbf{Characteristics} &\textbf{Bandwidth efficiency} &\textbf{Fairness} & \textbf{Complexity}&\textbf{Ref.} \\
\hline
\centering
Ordered power allocation & Allocates more power for weak users  & Moderate & Moderate & Low&\cite{Saito:2013,saito2013system,ding2014performance,yuanwei_JSAC_2015}\\
\hline
\centering
Max-min rate fairness& Absolutely fairness for users &Low & High& High& \cite{Timotheou:2015,Yuanwei2016NOMA,Cui2016NOMA} \\
\hline
\centering
Proportional fairness & Maximizes users' geometric mean &Moderate & Moderate& Low& \cite{Liu2015PIMRC,Liu2016ICC,Mei2016ICC,Otao2015fairness} \\
\hline
\centering
 $\alpha$ utility function& Imposes fairness & High  & Moderate &  Moderate&---\\
\hline
\centering
 Weighted sum rate& Achieves fairness with the aid of MAC layer & High  & Low & Low& \cite{sun2016optimal}\\
\hline
\centering
Jain's fairness index comparison & Only a fairness performance metric & ---  & --- & --- &\cite{al2014uplink,lei2016power,WPT_NOMA_2015}\\
\hline
\end{tabular}
\end{center}
\label{table:PA NOMA}
\end{table*}

\subsubsection{Cognitive radio inspired power control}
\begin{figure}[t!]
    \begin{center}
        \includegraphics[width=8cm]{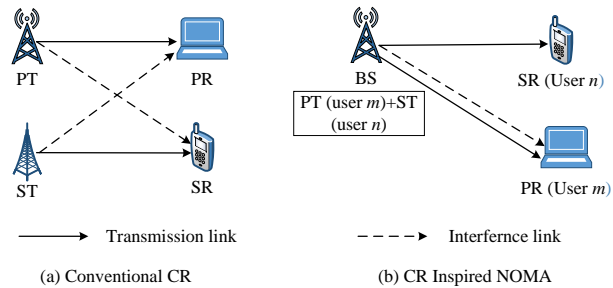}
     \caption{Comparison of convectional CR and CR inspired NOMA.}
            \label{CR inspired NOMA}
    \end{center}
\end{figure}

The objective of CR inspired power control relying on NOMA is to guarantee the QoS of weak users by constraining the power allocated to the strong user.
Inspired by the CR concept~\cite{goldsmith09}, NOMA can be regarded as a special case of CR networks~\cite{ding2014pairing,Yangzheng2016CRNOMA}. More specifically, still considering a downlink scenario supporting two
users, Fig. \ref{CR inspired NOMA} compares conventional CR and CR inspired NOMA. The BS can be viewed as the combination of a primary transmitter (PT) and a secondary transmitter (ST), which transmits the
superimposed signals. The strong user (User $n$) and the weak user (User $m$) can be regarded as a secondary receiver (SR) and a primary receiver (PR), respectively. By doing so,
the strong User $n$ becomes capable of accessing the spectrum occupied by the weak User $m$ under predetermined interference constraints, which is the key feature of the classic underlay CR.
The concept of CR-inspired PA in NOMA was proposed by Ding \emph{et al.} in~\cite{ding2014pairing}, who investigated the PA of user-pairing-based NOMA systems.

The key advantages of  cognitive PA are summarized as follows:
\begin{itemize}
  \item \textbf{Guaranteed QoS}: by applying cognitive PA, the QoS requirements of the weak user are guaranteed, which is especially vital in real-time safety-critical applications.
  \item \textbf{Fairness/throughput tradeoff}: cognitive PA is capable of  striking a beneficial tradeoff between the overall system throughput and the individual user fairness, where the targeted data rate of the weak user has to be satisfied by appropriate PA.
  \item \textbf{High flexibility}: cognitive PA offers a high degree of freedom for the BS to explore the opportunistic support of the strong user.
  \item \textbf{Low complexity}: compared to the optimal PA approach, cognitive PA imposes a lower complexity during PA. This becomes particularly  useful when the channel ordering and PA constraints are not convex and hence finding an appropriate PA scheme becomes a challenge, especially in  multiple antenna aided NOMA scenarios.
\end{itemize}

Motivated by its advantages mentioned above, the cognitive PA policy was invoked  for characterizing  MIMO-NOMA systems in~\cite{Zhiguo_general:2015,Zhiguo2016IoT}.
More particularly, in addition to investigating the convectional downlink cognitive PA conceived for MIMO-NOMA scenarios, the authors of~\cite{Zhiguo_general:2015} also designed a more sophisticated CR NOMA PA scheme for uplink MIMO-NOMA scenarios.
In~\cite{Zhiguo2016IoT}, in an effort to find a PA strategy suitable for SU-MIMO IoT scenarios,
a cognitive PA policy was designed for ensuring that SIC may indeed be carried out at the strong user. 

\subsection{User Scheduling in Dynamic Cluster/Pair Based Hybrid MA Networks}\label{multiple channel power allocation}
Due to the low-complexity design of NOMA and the new degree of freedom by the power dimension, it is widely accepted both by the academia and industry that NOMA constitutes a convenient
an ``add-on" technique for establishing spectral efficient hybrid MA networks. In an effort to design
hybrid MA networks, the users have to be assigned to different pairs/clusters. By doing so, how to pair/cluster the users becomes an interesting problem, which is valuable  to examine.
Several excellent contributions have researched the potential issues aiming for obtaining efficient user pairing/clustering strategies
with the objective of improving the performance of both a single pair and of the entire networks, as detailed below.
\subsubsection{A single user pair's performance}
User pairing is capable of beneficially reducing the complexity of the NOMA design, when establishing advanced hybrid MA networks. By focusing
on fully exploiting the potential gains of NOMA over conventional OMA from the perspective of a single user pair, Ding \emph{et al.}~\cite{ding2014pairing}
investigated both the sum rate  and the individual rates of the two users. It was demonstrated that a higher sum rate gain  can be attained by NOMA
over
OMA, when the channel conditions of the pair of NOMA users are rather different, which is line with our expectation in the context of SIC-aided detection.
In \cite{yuanwei_JSAC_2015}, the impact of spatial location on a pair of NOMA users
was examined by employing a stochastic geometry model, in conjunction with three different user selection schemes.
In \cite{Qin2016PLS_NOMA}, Qin \emph{et al.} examined the secrecy performance of a specifically selected user pair in large-scale networks and showed that the
secrecy diversity order of the user pair was determined by the user having the weaker channel. User pairing was invoked for MIMO-NOMA scenarios in \cite{Zhiguo_general:2015,Zhiguo2016IoT}. We note that the user paring techniques of~\cite{ding2014pairing,yuanwei_JSAC_2015,Qin2016PLS_NOMA,Zhiguo_general:2015,Zhiguo2016IoT} can be further improved by
 PA, which is hence a promising research direction.
\subsubsection{User clustering/pairing from the perspective of the overall system's performance}
Note that allocating NOMA users to different orthogonal RBs (e.g. sub-channels/sub-carries/clusters) generally turns out to be a  non-deterministic
polynomial-time (NP)-hard
problem. It is computationally prohibitive to obtain optimal results by performing an exhaustive search, when the number of users and RBs becomes
high, especially when the number of user/RBs is dynamically fluctuating.
Hence investigating efficient low-complexity user allocation algorithms, which are capable of achieving attractive performance-complexity trade-offs is necessary.
Some of the research contributions in the field of resource allocation can be found in~\cite{Lei2015GC,Di2015GC,Parida2014GC,Yang2015ICC,Liu2015PIMRC,Zhang2016ICC,Fang2016ICC,Mei2016ICC,Elsaadany2016ICC,Fu2016ICC,Han2016ICC,Hossain2016NOMA}.
Below we list some of the promising approaches as follows:
\begin{itemize}
  \item \textbf{Combinatorial relaxation:} A popular method of tackling this kind of issue is to relax the constraint of the combinatorial problem
  by using continuous variable of $\eta  \in \left[ {0,1} \right]$
instead of a binary variable of $\eta  \in \left\{ {0,1} \right\}$, where $\eta$ indicates whether a user is allocated to a RB block. By doing so, the original NP-hard
problem is transferred to a convex problem and can be solved by invoking the classic Lagrangian dual method~\cite{boyd2004convex}. However, this kind
of relaxation will result in a non-trivial duality performance gap between the original problem and the simplified one.
  \item \textbf{Monotonic optimization:} Due to the non-convex nature of many sophisticated resource allocation problems,
  finding the optimal solution using convex optimization theory is rather challenging. In addition to convexity, the monotonicity is another
  important  property, which can be exploited for efficiently solving non-convex optimization problems \cite{Jorswieck2010monotonic,qian2009monotonic}.
  Note that the resource allocation problem of NOMA scenarios is usually non-convex due to its inter-user interference in the same RB.
  In this context, Sun \emph{et al.}~\cite{sun2016optimal} invoked the classic monotonic optimization approach for developing an optimal solution
  for their joint power and subcarrier allocation problem. A low complexity suboptimal approach was also proposed for striking a performance-vs-complexity tradeoff.
  \item \textbf{Matching theory:}  Matching theory, as a powerful mathematical modeling tool conceived for solving the combinatorial user allocation problems, is capable of overcoming
  some of the shortcomings
  imposed by the widely used game theory, such as the distributed-limited implementation and unilateral equilibrium deviation~\cite{Gu2015Matching}.
  As such, matching theory has recently emerged
  as a promising technique of tackling the resource allocation problem in wireless networks, also in NOMA scenarios \cite{Di2015GC,zhang2016sub,Jingjing2016D2D,Jingjing2017HetNets}.
  More particularly, Di \emph{et al.}~\cite{Di2015GC} invoked many-to-many
  two-sided matching theory for resource allocation in the downlink of NOMA systems. A potential shortcoming of matching theory is that both the
  users and resources should have a predefined preference list. However, these preference lists may vary as the channel conditions fluctuate, which
  hence requires further research.
  \item \textbf{Heuristic algorithms:} Heuristic algorithms are commonly used for solving computationally complex problems.
   The core idea of heuristic algorithms is to obtain approximate solutions for the original optimization problem at an
  acceptable computational complexity. Inspired by this, many heuristic algorithms such as meta-heuristics~\cite{Yang2015ICC}, greedy techniques~\cite{Parida2014GC}
   and others~\cite{Liu2015PIMRC}
   have been
  adopted by researches in NOMA scenarios for allocating resources.  Given the diverse nature of heuristic algorithms~\cite{kokash2005introduction},
 they may lead to significantly different performance vs complexity trade-offs.
\end{itemize}

It is worth pointing out that apart from applying intelligent user scheduling strategies,  optimizing PA for each RB/user
is capable of further improving the attainable network performance. However, the joint optimization of user scheduling and PA is a non-trivial problem.
Except for \cite{sun2016optimal}, which jointly considers the PA as well as user association and obtains the optimal solution,
the commonly-adopted
approaches rely on decoupling this correlated problem into a pair of sub-problems~\cite{Di2015GC,Yang2015ICC,Parida2014GC}.
 As such, the aforementioned PA and user scheduling algorithms can be invoked for finding a  sub-optimal
solution. However, the existing literature of resource allocation is predominantly focus on maximizing the performance attained, whilst there is
a distinct lack of a systematic performance vs computational complexity analysis. Motivated by filling this gap,
Lei \emph{et al.}~\cite{lei2016power} characterized the tractability of NOMA resource allocation problems under several practical constraints.
A combined Lagrangian duality and dynamic programming  algorithm was also proposed in~\cite{lei2016power} for solving the joint optimization
of the PA and user scheduling problem of NOMA networks.

\begin{table*}[htbp]
\caption{Summary of resource control solutions for NOMA}
\begin{center}
\centering
\begin{tabular}{|l|l|l|l|}
\hline
\centering
 \textbf{UA approaches} &\textbf{Advantages} &\textbf{Disadvantages} & \textbf{Ref.} \\
\hline
\centering
Combinatorial relaxation & Convert to convex problem & Existence of duality gap&  --- \\
\hline
\centering
Monotonic optimization & Optimal solution & Relatively high complexity&  \cite{sun2016optimal} \\
\hline
\centering
Matching theory & Achieve near-optimal performance & Requires predefined preference list &  \cite{Di2015GC,Zhang2016ICC,Fang2016ICC,Jingjing2016D2D}\\
\hline
\centering
Heuristic algorithms & Flexible complexity-performance tradeoff  & Unstable performance &  \cite{Parida2014GC,Yang2015ICC,Liu2015PIMRC}\\
\hline
\end{tabular}
\end{center}
\label{table:PA NOMA}
\end{table*}

\subsection{Software-Defined NOMA Network Architecture}
Inspired by the concepts of the emerging SDN paradigm~\cite{kreutz2015software}, we propose the novel SD-NOMA concept,
which provides the desired degree of flexibility for controlling resources.  By decomposing the resource allocation and control problem into
complex tractable problems,
the SD-NOMA concept makes it much easier to create new abstractions in terms of power optimization, interference management,
user association and dynamic user clustering/pairing. Before introducing the proposed SD-NOMA strategy, we first review the compelling concept of
SDR. The essential idea behind SDR is to implement the baseband signal processing algorithms in software instead of
hardware,
which hence exhibit a high grade of flexibility and reconfigurability, when  designing agile instantaneously adaptive  communication systems.
In this spirit, a practical open source SDR-based NOMA prototype has been designed for a typical  two-user downlink scenario by Xiong \emph{et al.}~\cite{Xiong2015SDR},
which bridges the theoretical and practical aspects of NOMA. Both the hardware and software architecture of the NOMA prototype systems designed were elaborated on.
The link-level simulation results showed that the performance of the SIC procedure was significantly influenced both by the PA as well as by the modulation scheme employed.

\begin{figure}[t!]
    \begin{center}
        \includegraphics[width=8cm]{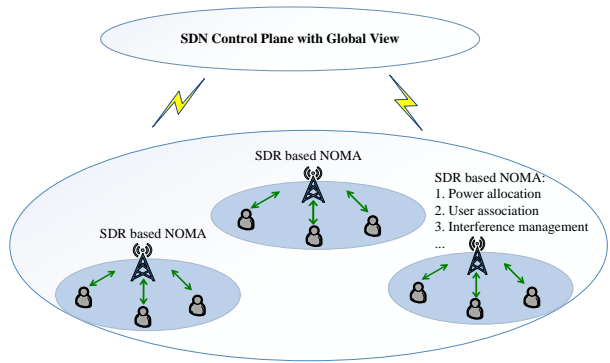}
     \caption{Illustration of Software-Defined NOMA Network Architecture.}
            \label{SD_NOMA}
    \end{center}
\end{figure}

In contrast to the SDR-based NOMA, the key idea of our proposed SD-NOMA network architecture is that the SDN controller has a global view of both
the available resources and of the tele-traffic across the whole network, as shown in Fig. \ref{SD_NOMA}. However, each component
may still rely on the SDR-based NOMA concept. In other words, the SD-NOMA network architecture can globally control the SDR-based NOMA components via a central  SDN controller.
 Considering the user association and PA in the context of a cluster-based MIMO-NOMA scheme \cite{ding2015mimo} as an example, the SDN
 controller becomes capable of
  associating the different users with a potential cluster and allocating the most appropriate power levels to different users,
  whilst at the same time considering the interference arriving not only from the users in the same cluster but also from other clusters. This approach formulates
 the PA of NOMA as a global optimization problem, since the SDN controller is aware of the entire network's state. Therefore, SD-NOMA
constitutes an attractive 5G network structure.
\subsection{Discussions and Outlook}
The distinctive co-channel interference characteristics of NOMA usually lead to  a non-convex optimization problem.
The aforementioned research contributions have provided numerous beneficial techniques of dealing with the resource control problem. However,
for complexity reasons,  most of these techniques opted for decoupling the user scheduling and PA into consecutive subproblems,
which usually result in sub-optimal solutions.
The problem of finding optimal solutions by jointly considering both the user scheduling and PA is still far from being well understood.
Hence advanced optimization techniques requiring low-complexity algorithms are required for solving this problem.

In summary, the design of attractive NOMA solutions hinges on finding the Pareto-optimal solutions, as discussed for example
in an ad hoc networking context in \cite{Alanis2017Access}, where none of the system characteristics can be improved without degrading at least
one of the others. In the interactive 7-mode networking demo a three-component objective function constituted by the BER,
the power and the delay was animated.



\section{Compatibility of NOMA with Other Technologies towards 5G and Beyond}
As one promising technique in future 5G networks, one of the main challenges of NOMA is that how well it is to be compatible with other emerging techniques
for meeting the requirements of 5G.
In this section, we survey the existing research contributions considering the co-existence of NOMA with other 5G proposals.

\subsection{NOMA in HetNets}
Dense HetNets, as one of the hot 5G technologies~\cite{andrews2014will},  are capable of significantly improving the network capacity.
The core idea of HetNets is essentially to move the low-power BSs closer to the served users in order to form small cells under the over-sailing macro cells.
However, due to the co-channel nature of macro cells and small cells, the users surfer both from inter-layer interference as well as from intra-layer
interference. To intelligently cope with the interference arriving from the other co-channel layers, in~\cite{Xu2015GC} Xu \emph{et al.} proposed a cooperative NOMA scheme
for HetNets and minimized  the inter-user interference with the aid of DPC precoding. The effect of distinctive power disparity between the macro BSs and pico BSs was investigated.

\begin{figure}[t!]
    \begin{center}
        \includegraphics[width=3.5in]{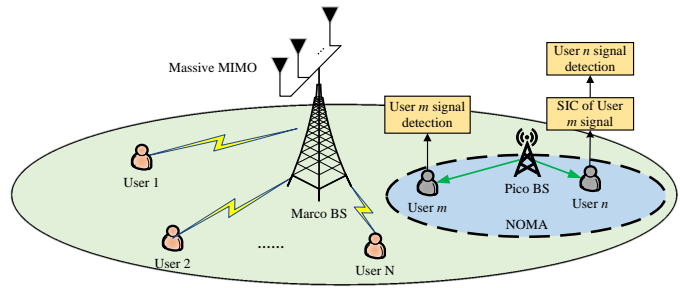}
        \caption{Illustration of NOMA and massive MIMO based hybrid HetNets.}
        \label{System model NOMA HetNets}
    \end{center}
\end{figure}

Additionally, as a benefit of aiming to take full advantage of both massive MIMO and HetNets techniques including high array gains, reliable BS-user links etc.,
massive MIMO aided HetNets solutions are promising for the emerging 5G systems \cite{Adhikary2015Hetnets}. The basic philosophy of
massive MIMO assisted HetNets is to install hundreds/thousands of antennas at the macro BSs for offering an unprecedented level of spatial degrees of
freedom, whilst using
a single antenna at the densely positioned small-cell BSs. In an effort to further enhance the bandwidth efficiency of small cells, a promising NOMA aided
and massive MIMO based framework was proposed in \cite{Yuanwei2016HetNets}, where a massive MIMO  system was adopted by
the macro cells to simultaneously serve $N$ users and then user pairing based NOMA transmissions were adopted
by the small cells, as shown in Fig. \ref{System model NOMA HetNets}. More specifically, stochastic geometry was invoked for modelling the  $K$-tier
HetNets considered and to analyze the efficiency attained. The key feature of this framework is that it integrates the potential
advantages of both NOMA (e.g., high bandwidth efficiency, faireness/throughput tradeoff, etc.) and of HetNets (low power consumption, spatial spectrum reuse, etc.)
whilst simultaneously  relying on a sophisticated cluster based MIMO-NOMA precoder design~\cite{higuchi2013non,ding2015mimo,Jinho:2015,Fainan2015TSP,Zhiguo_general:2015,Zhiguo2016IoT}.

\subsection{NOMA in Millimeter Wave Communications}

MmWave communications have been recognized as a promising technique in 5G networks and beyond due to their large bandwidths in the high-frequency spectrum \cite{pi2011mmWave}. The severe propagation path loss of mmWave channels and their low penetration capabilities \cite{Alkhateeb2014mmWave} necessitates the redesign of MA techniques, especially when aiming to support massive connectivity in dense networks, where hundreds/thousands of users have to be served within a small area. NOMA can be regarded as a potent MA technique capable of co-existing with mmWave networks due to the following reasons:
\begin{itemize}
  \item The highly directional beams used in mmWave communications lead to correlated channels, which may degrade the performance of convectional OMA systems, but they are amenable to NOMA.
  \item The sharp beams of mmWave networks effectively suppress the inter-beam interference among users, which is suitable for supporting NOMA in each beam.
  \item The application of NOMA in mmWave is capable of enhancing the bandwidth efficiency and supporting massive connectivity.
\end{itemize}

\begin{figure}[t!]
    \begin{center}
        \includegraphics[width=3.5in]{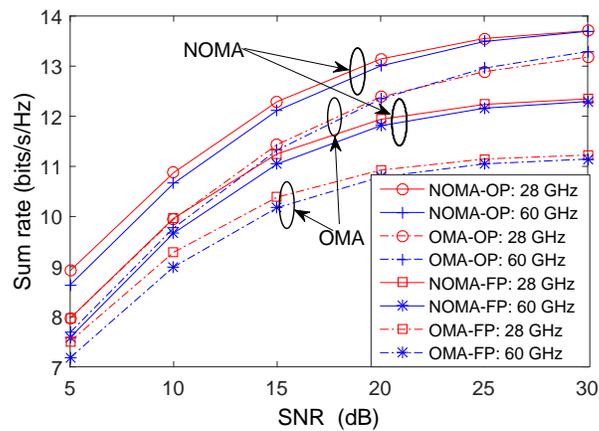}
        \caption{Illustration of the superiority of NOMA-mmWave over OMA-mmWave networks both at 28 GHz and 60GHz, ``OP" refers to optimized power allocation, ``FP" refers to fixed power allocation. The detailed parameter settings are found in \cite{cui2017optimal}.}
        \label{NOMA mmwave}
    \end{center}
\end{figure}

Inspired by this, some initial research contributions examining NOMA in mmWave networks have been disseminated in \cite{Zhiguo2016mmWave,cui2017optimal}. In \cite{Zhiguo2016mmWave}, Ding \emph{et al.} investigated the co-existence of NOMA and mmWave
solutions relying on random beamforming. Due to the potential line-of-sight (LOS) blockages of mmWave systems, a thinning process aided stochastic geometry
model was used to evaluate the performance. As a further advance, Cui \emph{et al.} \cite{cui2017optimal} investigated the performance of NOMA-mmWave networks relying on partial CSI feedback. More particularly, the user scheduling and PA were jointly considered with the aid of matching theory and Branch-and-Bound approaches. Fig. \ref{NOMA mmwave} illustrates the sum rate of NOMA-mmWave versus the SNR at different frequencies. It is demonstrated that the proposed NOMA-mmWave system is capable of outperforming conventional OMA-mmWave systems.

\subsection{NOMA and Cognitive Radio Networks}
The 2010s have witnessed the rapidly increasing penetration of mobile devices (e.g., smart phones, tablets and laptops) all over the world,
which gave rise to increasing demand for spectral resources. As reported by the Federal Communications Commission (FCC), there are significant
temporal and spatial variations in the exploitation of the allocated spectrum. Given this fact, the CR concept---a terminology coined
by Mitola\cite{mitola2000cognitive}---inspired the community to mitigate the spectrum scarcity problem. The basic concept of CR is that at a certain
time of the day or in a geographic region,  the unlicensed secondary users (SUs) are allowed to opportunistically access the licensed spectrum  of primary users (PUs).
These CR techniques may be categorized
into
the interweave, overlay, and underlay paradigms~\cite{goldsmith09}:

\begin{itemize}
  \item \textbf{Interweave}: The interweave CR can be regarded as an interference avoidance paradigm, where
the SUs are required to sense the temporary slivers of the space-frequency domain of PUs before they access the channels~\cite{Haykin2005JSAC,Liang2008CR,Qin2016TWC,Qin2016TSP}.
The concurrent transmission of SUs and PUs is not allowed under the interweave paradigm.
  \item  \textbf{Overlay}: The overlay paradigm essentially constitutes an interference
mitigation technique. With the aid of the classic dirty paper encoding technique, overlay CR  ensures that a cognitive user becomes capable of transmitting
 simultaneously with a non-cognitive PU~\cite{goldsmith09}.
Additionally, SUs are capable of  forwarding the information of PUs' to the PU receivers, whilst superimposing their own signals as a reward for their relaying services.
  \item \textbf{Underlay}: The underlay CR operates like an intelligent interference control paradigm, where the SUs are permitted to access the spectrum allocated to PUs as
long as the interference power constraint at the PUs is satisfied~\cite{Zhao2007SPM}.
\end{itemize}

Table \ref{table:CR} provides a summary of the interweave, overlay, and underlay paradigms and briefly illustrates the differences among them.

\begin{table*}[htbp]\tiny
\caption{Summary and comparison of the three main CR paradigms}
\begin{center}
\centering
\begin{tabular}{|l|l|l|l|l|l|}
\hline
\centering
 \textbf{Paradigm} &\textbf{Concurrent Transmission} & \textbf{Constraints at SU} & \textbf{Interference Management}& \textbf{Complexity} & \textbf{NOMA-CR}   \\
\hline
\centering
 Underlay&  Not allowed & Limited by the interference power constraint & Interference avoiding & Low & \cite{Liu2016TVT,ding2014pairing}\\
\hline
\centering
Overlay & Allowed& No power constraints & Interference mitigating &  High & ---\\
\hline
\centering
 Interweave& Allowed& Limited by the range of sensing hole & Interference controlling & Medium & ---\\
\hline
\end{tabular}
\end{center}
\label{table:CR}
\end{table*}

One of the core challenges in both CR and NOMA networks is the interference management, whilst improving the bandwidth efficiency.
Hence it is natural to  link them for achieving an improved bandwidth efficiency.
Liu \emph{et al.} investigated the application of NOMA in large-scale underlay CR networks by relying on the stochastic geometry model~\cite{Liu2016TVT}.
The diversity order of the NOMA users was characterized analytically in two scenarios. The classic OMA-based underlay CR was also used as a benchmark to show the benefits
of the proposed CR-NOMA scheme. As mentioned in Section VI, Ding \emph{et al.}~\cite{ding2014pairing} proposed a novel PA policy for
NOMA, namely the CR-inspired NOMA PA, which constitutes a beneficial amalgam of NOMA and underlay CR.

To the best of our knowledge, CR-NOMA studies only exist in the context of the underlay CR paradigm. Hence both the interweave and overlay CR paradigms
have to be investigated in NOMA networks. It is worth pointing out that a significant research challenge of NOMA is to dynamically cluster/pair the NOMA users first,
followed by dynamically allocating the clusters/pairs to different orthogonal sub-channels. In the context of the interweave paradigm, intelligent
 sensing
has to be applied first, followed by user clusering/pairing of NOMA users, depending on the specific channel conditions sensed.

\subsection{NOMA-Based Device-to-Device Communication}
Due to the recent rapid increase in the demand for local area services under the umbrella of cellular networks,
an emerging technique, namely device-to-device (D2D) communication, may be invoked for supporting direct communications amongst devices without the assistance
of cellular BSs~\cite{fodor2012design}. The main advantages of integrating  D2D communications into cellular networks are: i) low-power support of
proximity services for improving the energy efficiency; ii) reusing the frequency of the over-sailing cellular networks in an effort to increase the bandwidth efficiency;
and iii) the potential to facilitate new types of peer-to-peer (P2P) services~\cite{Ma2015D2D}.

Note that one of the common features of both D2D and NOMA is that of enhancing the bandwidth efficiency by managing the interference among users within each RB. Motivated by this, it is desirable to invoke intelligent joint interference management approaches for fully exploiting the
potential benefits of both D2D and NOMA. In \cite{Jingjing2016D2D}, Zhao \emph{et al.} designed a novel NOMA-based D2D communication scheme, where several
 D2D groups were permitted to share the same RB with the cellular users. In contrast to the conventional D2D pair's transmission, the novel
 ``D2D group" concept was introduced, where a D2D transmitter was able to simultaneously communicate  with multiple D2D receivers  with the aid of NOMA.
  It was demonstrated that the proposed  NOMA-based D2D scheme is capable of delivering higher throughput than  conventional D2D communication.

\subsection{Discussions and Outlook}
The aforementioned techniques are capable of improving the bandwidth efficiency by exploiting NOMA, but they will also pose some challenges.
For example, adopting NOMA in HetNets, CR networks and D2D
communication scenarios will impose increased co-channel interference on the existing networks. Hence intelligent interference management
is desired.

Apart from the above-mentioned solutions, there are other emerging techniques, such as
C-RAN, full-duplex, and other solutions. As the extension of \cite{sun2016optimal},    Sun \emph{et al.} \cite{sun2016jounal} studied the multiple carrier
NOMA resource allocation problem for the scenario, where  a full-duplex BS supported several half-duplex users.  While the current research contributions
have laid a solid foundation, numerous open questions have to be resolved.

\section{Implementation Challenges and Standardization of NOMA}
Although NOMA have been recognized as a promising candidate for 5G and beyond, there are still several implementation challenges to be tackled. In this section, we identify some of the implementation issues and point out some potential approaches to solve them. Moreover, the standardization progress of NOMA will also be discussed in this section to show how NOMA paves the way to 5G and beyond.
\subsection{Error Propagation in SIC}
SIC is the key technique of user detection in NOMA systems. Nevertheless, a main drawback of implementing SIC is the inter-user error propagation issue, which propagates on as from one user to another, because a decision error results in subtracting the wrong remodulated signal from the composite multiuser signal, hence resulting in residual interference \cite{verdu1998multiuser,Duel1995CDMA}. More specifically, if the messages of previous users are not correctly decoded, the reconstruction of those signals will result in flawed decoding of the remaining users' messages, which leads the accumulation of the decoding errors. Most existing research contributions in the context NOMA are based on the assumption that the SIC receivers are capable of perfectly cancelling the interference. Actually this assumption cannot be readily satisfied in practice due to the inaccurate PA and imperfect channel decoding. Several researchers have recognized the opacity of error propagation issues and investigated the effect of imperfect SIC on uplink NOMA \cite{tabassum2016modeling} systems, on full-duplex NOMA networks \cite{Xinwei2017RS} and on C-RAN NOMA networks \cite{Jingjing2017CRAN}.

Some recent research contributions have identified several possible techniques for solving the error propagation issues. The first one is to use multistage channel estimation to improve the SIC \cite{Kobayashi2001JSAC} at the cost of increasing the complexity. The second possible approach is to apply iterative SIC aided receiver to overcome the error propagation effects, as proposed by Zhang and Hanzo \cite{Zhang2011CST}. Another strategy is to take the channel estimation error into considerations when designing the power control algorithms, which was proposed by Buehrer \cite{Buehrer2001TCOM}. More particularly, more power is allocated to the users who will carry out SIC later for compensating the residual interferences. Nevertheless, invoking this approach imposes a capacity payoff \cite{Andrews2005SIC}.

\subsection{Channel Estimation Error and Complexity for NOMA}
Channel estimation plays a more significant role in NOMA than in OMA systems, since the channel estimation errors will result in  ambiguous user ordering as well as inaccurate power control, which will in turn affect the SIC decoding accuracy. Naturally, perfect CSI cannot obtained in practice. High performance near-optimal conventional channel estimation algorithms impose an unacceptable system overhead and computational complexity on NOMA systems, especially for MIMO-NOMA scenarios. In order to solve those problems, researchers dedicated their efforts mainly to the following three aspects.

The first approach is to propose more effective channel estimation designs for striking  a good complexity-performance tradeoff. As mentioned in Section III, CS can be applied as an effective technique of reducing the complexity \cite{Rao2014TSP}. The second approach is to rely on partial CSI. This is motivated by the fact that the large scale fading fluctuates less rapidly than the small scale fading. As mentioned in Section II, the use of partial CSI in downlink NOMA was studied in~\cite{Yang2016Tcom,shi2015outage,Cui2016NOMA}. The third approach is to use limited feedback for reducing the overhead \cite{xu2016outage}. Finally, iterative joint decision-directed channel estimation and data detection can be invoked as detailed in \cite{hanzo2010mimo}.

\subsection{Security Provisioning for NOMA}
Security issues are of great significance in each generation of networks. The broadcast nature of the wireless medium makes it vulnerable  to  eavesdropping. Physical layer security (PLS), which was proposed by Wyner as
early as 1975~\cite{wyner1975wire}, has become an appealing technique of improving the confidentiality of wireless communications. In contrast to the
traditional approaches,
which design cryptographic protocols in the upper layers, PLS aims for exploiting the specific characteristics of wireless channels in the physical layer for transmitting confidential messages.
The key idea of achieving perfect secrecy in wiretap channels is to ensure that the capacity of the desired channel is higher than that of the eavesdropper's channel.
Triggered by the rapid development of wireless networks, PLS has been considered  in diverse scenarios~\cite{Mukherjee:2011,ZhiguoDing2012,Yuanwei:2015,zou2014security,liu2015PLS}.

Motivated by the security concerns of wireless communications,  PLS measures have also been proposed for NOMA networks in order to combat eavesdropping~\cite{Qin2016PLS_NOMA,Yuanwei2017TWC,Zhang2016PLS_NOMA,Yuanwei2017TWC,he2016design,ding2016spectral}.
In~\cite{Qin2016PLS_NOMA}, Qin \emph{et al.} examined the PLS of single-antenna NOMA in large-scale networks by invoking stochastic geometry,
 where the BS communicates with randomly distributed NOMA users. A protected zone may be adopted around the BS, where the intended users benefit from a high capacity in
 order to establish an eavesdropper-exclusion
 area to enhance the PLS with the aid of careful channel-ordering of the NOMA users. As a further development, Liu \emph{et al.} \cite{Yuanwei2017TWC} investigated the PLS of multiple-antenna
 NOMA scenarios, where artificial noise was generated at the BS for degrading the channels of eavesdroppers. Zhang \emph{et al.}~\cite{Zhang2016PLS_NOMA} investigated the PLS of
 SISO-NOMA networks, where the secrecy sum rate was maximized and the optimal PA was characterized in closed-form
 expressions.

\subsection{Maintaining the Sustainability of NOMA with RF Wireless Power Transfer}
One of the key objectives of future 5G networks is to maximize the energy efficiency and to support energy-constrained wireless devices. Energy consumption is a critical factor in maintaining the sustainability
of wireless networks, especially for the devices, which invoke a  high cost of replacing the batteries.
EH is an effective technique of extending the battery recharge period. Hence it has recently received remarkable attention~\cite{zhang2013mimo,nasir2013relaying,Liu2014two_way,Qin2015WPT,liu2015secure,Liu2015WPT}.
However, the traditional EH techniques relying on solar-wind-vibration-and thermoelectric-effects, which depend on the location,
environment, time of the day, etc.  In contrast to the conventional EH techniques, radio-frequency (RF) wireless power transfer (WPT) provides a more
flexible approach for powering energy-constrained devices. Another motivation behind this approach lies in the fact that most devices are surrounded by
ubiquitous RF signals. As a consequence, even interfering signals can be regarded as the potential EH sources.

To elaborate, NOMA relying on RF WPT techniques in wireless networks has been studied in~\cite{liu2015cooperative,yuanwei_JSAC_2015,WPT_NOMA_2015,sun2016transceiver}.
More particularly, in \cite{liu2015cooperative} the application of WPT to NOMA networks was investigated, where the users are randomly located.
As a further development, in \cite{yuanwei_JSAC_2015}, a new cooperative simultaneous wireless information and power transfer (SWIPT) based NOMA protocol was proposed.
In order to elaborate on the impact
 of user association, three user selection schemes based on the user distances from the base station were proposed in \cite{yuanwei_JSAC_2015}.
 The  analytical results of \cite{yuanwei_JSAC_2015} confirmed that invoking SWIPT techniques did not degrade the diversity gain compared to
that of conventional NOMA.  In \cite{WPT_NOMA_2015}, the uplink of NOMA transmission was considered in the context of energy constrained users,
 who can harvest energy from the
 BS by adopting the ``\emph{harvest-then-transmit}" protocol. The results demonstrated that NOMA is also capable of providing considerable
  throughput, fairness and energy-efficiency improvements.

\subsection{State-of-the-art for Standardization of NOMA}

NOMA has  recently been included into LTE-A, terms MUST \cite{MUST2016ICC}. More specifically, at the 3GPP meeting in May 2015, it was decided to include MUST into LTE Advanced. Afterwards, at the 3GPP meeting in August 2015, 15 different forms of MUST have been proposed by Huawei, Qualcomm, NTT DOCOMO, Nokia, Intel, LG Electronics, Samsung, ZTE, Alcatel Lucent, etc. Finally, at the 3GPP meeting in December 2015, NOMA has been included into LTE Release 13 \cite{3GPP}. It is worth noting that the MUST technique may be made compatible with the existing LTE structure. In other words, NOMA allows two users to be served at the same OFDM subcarrier without changing the current structure. Various non-orthogonal transmission schemes have been proposed for the MUST items \cite{3GPP,MUST2016ICC}, which can be generally classified into three categories \cite{3GPPR1}, namely, 1) superposition transmission with an adaptive power ratio on each component constellation and non-Gray-mapped composite constellation; 2) Superposition transmission with an adaptive power ratio on component constellations and Gray-mapped composite constellation;  3) Superposition transmission with a label-bit assignment on composite constellation and Gray-mapped composite constellation. The examples of transmitter processing candidate can be found in Figure 6 of \cite{Zhiguo2015Mag}.  The  NOMA-like MUST architecture works as follows. The coded bits of near users and far users are first separately input in a bit converter then modulated with the aid of quadrature amplitude modulation (QAM) or quadrature phase shift keying mapper (QPSK). The modulated symbols are superposed with allocating appropriate powers to transmit.  Actually, the first category can be regarded as a special case of the second category, since coded bits in the first category are modulated directly to mappers without inputting into the bit converter.  Both of these two categories are capable of supporting flexible power partition among users \cite{Yuan2016Mag}.  For the third category, it is a bit-partition-based scheme which is in contrast to the power partition scheme adopted in the first and second categories. Regarding the detailed differences among the mentioned three schemes, interested readers may refer \cite{Zhiguo2015Mag,Yuan2016Mag} to identify the transmit structures of the three schemes. Regarding receivers, interference cancelation is required to carried out. The interference scenarios are different between network assisted interference cancellation system (NAICS) and MUST since the receivers of NAICS and MUST are employed by different groups of users. More particularly, for a two-user case which consists of a cell-edge user and a cell-center user, NAICS receivers are typically used for cell-edge users while MUST is typically used for cell-center users. By doing so, the network performance can be enhanced.

In addition to LTE-A, NOMA had also been included in the forthcoming digital TV standard, by the Advanced Television Systems Committee (ATSC) 3.0 \cite{Zhang2016LDM}, termed as layered division multiplexing (LDM) for providing significant improvements in terms of service reliability, system flexibility, and spectrum efficiency. This standard will generate significant impact on digital TV industry. Moreover, in the white paper of  NTT DOCOMO, NOMA has been identified as a key technique for 5G. The system-level performance of NOMA has also been demonstrated by NTT DOCOMO\cite{benjebbour2013system}. It is worth pointing out that the key challenges of implementing NOMA in industry is the decoding complexity increases at receivers as the number of users increase. Another potential challenge is that the security and privacy of far users should be protected at near user side due to the characteristic of SIC, which may depend on key generations from upper layer. Although NOMA is reviewed as a promising candidate for 5G and beyond, there are various forms of NOMA (which will be detailed in the following section). The standardization process for NOMA is still ongoing.

\subsection{Discussions and Outlook}

In addition to the aforementioned implementation issues, there are also other imperfections to be dealt with for NOMA. Since NOMA relies on MA at different power levels, the strength of the received signals is deliberately different, which bring about new challenges for accurate analog-to-digital (A/D) conversion. On the one hand, for strong signals, a large voltage range is needed. On the other hand, for weak signals, high-resolution ADCs are required for supporting accurate quantization at small levels. In practice, considering the cost and system complexity, it is impossible to apply ADCs, which have both in large voltage range as well as a high resolution. The quantization errors are unavoidable. It is important to seek a good performance-complexity tradeoff.

Accurate synchronization is another significant issue in NOMA, which is to a degree neglected in the existing literature. Actually, perfect synchronous transmissions cannot be achieved in practice due to the dynamic mobile environment of users, especially for uplink NOMA transmission.  In order to solve the synchronization issues, two possible approaches can be adopted. The first one is to propose accurate pilot design for reducing the time synchronization errors. The second approach is to investigate novel asynchronous communication schemes. Haci \emph{et al.} \cite{Haci2017NOMA} proposed a new IC scheme for asynchronous NOMA aided OFDM systems. It was demonstrated that the system performance largely depends on the relative time offset among the interfering users. Other imperfections such as the impact of filtering distortion between transmitters and receivers on the performance degradation of NOMA systems are still unknown, which is a promising future direction to consider.

\section{Practical Forms of NOMA}\label{other NOMA techniques}
In the previous sections, power-domain NOMA has been investigated from diverse perspectives, including multi-antenna techniques,
CR NOMA, the resource management problems of NOMA and the co-existence of NOMA as well as other 5G solutions. In practice, NOMA can
be implemented in various forms, such as code-domain and power-domain NOMA, as summarized in this section. We may classify the practical forms of NOMA also into single-carrier and multiple-carrier NOMA. Their advantages and disadvantages will be discussed in this section.

\subsection{Single-Carrier NOMA}
We commence our discussions from the single-carrier NOMA design, which will lay the  foundations for the multi-carrier NOMA design in the next subsection.
\subsubsection{IDMA}
The key idea of the IDMA scheme relies on a unique user-specific chip-interleaver for distinguishing the signals of different users~\cite{IDMA2006liping}.
Hence IDMA may be viewed as chip-interleaved CDMA, which has the benefit of a high diversity gain, because if one or two chips are corrupted, the corresponding
spreading sequence could still be recovered with the aid of low-complexity chip-by-chip iterative multi-user detection (IMD) strategy. A comprehensive
comparison of IDMA and CDMA was presented in \cite{IDMA2012comparision} in terms of its performance vs complexity.
\subsubsection{LDS-CDMA}
LDS based CDMA constitutes an enhanced version of CDMA \cite{LDSCDMA2008Hoshyar}, which is inspired by the investigations
on the low-density parity check (LDPC) codes \cite{LDPC1962Gallager}. For conventional CDMA, one possible solution is to assign orthogonal spreading
sequences for each user, and hence low-complexity receiver can be invoked at receivers for eliminating interference. Nonetheless, this orthogonal
spreading code design is only capable of supporting equal number of users and chips, which motivates the development of non-orthogonal spreading code
design. So-called sparse spreading sequences are employed instead of the dense
spreading sequences of conventional CDMA, such as orthogonal variable spreading factor (OVSF) codes. However, such design requires sophisticated multi-user decoding at receivers. For LDS-CDMA, a message passing algorithm (MPA)
based the muti-user detection technique can be applied at
the receiver, which is capable of  achieving the near maximum likelihood (ML) detection performance.
\subsubsection{LPMA}
Lattice partition based multiple access (LPMA) is a new downlink non-orthogonal multi-user superposition transmission scheme, which was proposed by
Fang \emph{et al.} \cite{fang2016lattice}. LPMA achieves a beneficial multiplexing gain both in the power domain and in the code domain.
More explicitly, power-domain multiplexing has the potential of increasing the throughput by superimposing different-power streams. By contrast,
code-domain multiplexing superimposes several streams by exploiting that the linear combination of lattice codes also results in a lattice code.
More specifically, LPMA encodes the information of users by applying lattice coding at the transmitters and invokes SIC at the receivers for detection.
It is worth pointing out that a specific advantage of LPMA is that it has the potential of circumventing a specific impediment of power-domain NOMA,
namely that the performance gain attained relies on the channel quality difference of users. However, LPMA imposes a higher
encoding and decoding complexity than power-domain NOMA.

\subsection{Multi-Carrier NOMA}
Given the remarkable advantages of OFDM, this mature wave form design is likely to be incorporated in 5G networks. Explicitly, one of the main benefits of multi-carrier
techniques is that instead of the short dispersion-sensitive symbols of serial modems, many parallel long-duration
dispersion-resistant symbols are transmitted over dispersive channels. Hence these long-duration symbols are only mildly affected by the same
CIR and can be readily equalized by a single-tap frequency-domain equalizer.
The NOMA techniques are expected to coexist with OFDM. In this subsections, several forms of multi-carrier NOMA designs are discussed.
\subsubsection{LDS-OFDM}
LDS-OFDM \cite{LDSOFDM2010Hoshyar} is essentially a combination of LDS-CDMA and OFDM, which can be regarded as  an advanced variant of LDS-CDMA
in a multi-carrier form. Therefore, the same sparse spreading sequences  are applied at the transmitters and the same MPA based detection is
adopted at the receivers. The data-stream mapping process consists of two steps: 1) Each bit of the data streams is first spread by the low density
spreading sequences; and 2) The data streams are then transmitted over different subcarriers by applying an OFDM modulator. There are several
techniques of mapping the spreading sequences to OFDM, each having different pros and cons. For example, each chip
may be mapped to a different subcarrier for achieving frequency-domain diversity. Alternatively, each subcarrier may convey a spread low-rate stream.
However, this design results in a higher complexity due to the application of the MPA detection compared to a conventional OFDMA design.
\subsubsection{SCMA}
SCMA \cite{Nikopour2013SCMA} also relies on sparse spreading codes for ensuring that each RB can support more than one user with the aid of low density
 spreading, hence resulting in a scheme similar to LDS-CDMA and LDS-OFDM. Considering an example of six users and four sub-carriers, a typical signature
 matrix can be illustrated as follows:

\begin{align}\label{SCMA}
{\bf{S}} = \left[ {\begin{array}{*{20}{c}}
1&0&0&0&1&0\\
1&0&1&1&0&1\\
0&1&1&0&1&1\\
0&1&0&1&0&0
\end{array}} \right].
\end{align}

It is worth noting that for $\bf{S}$ in \eqref{SCMA}, each column has only two non-zero entries, which implies that each user is only allowed to
occupy two sub-carriers. This is one of the key characteristics of SCMA. In addition to these sparse spreading sequences, SCMA also relies on
multi-dimensional constellations for generating its codebooks in order to achieve a so-called constellation shaping gain. More particularly, SCMA
allows the employment of fewer constellation points to be used at a given throughput. For example, the four-level quadrature amplitude modulation (4-QAM)
constellation can be reduced to a smaller constellation, because some bits can be conveyed in the sparse-code domain.
However, this requires sophisticated codebook design \cite{Taherzadeh2014SCMA,Bao2016SCMA,Peng2017SCMA}. SCMA relies on the joint encoding
of the signals of multiple sub-carriers, hence joint decoding is required at the receivers. For this large joint alphabet MPA based
detection is capable of achieving near-ML performance at a fraction of the ML complexity.

\subsubsection{PDMA}
PDMA is another form of sparse signature matrix based multi-carrier NOMA \cite{ChenPattern7526461}. Again, we use the previous example of six users
and four subcarriers relying on the pattern matrix of:

\begin{align}\label{PDMA}
{\bf{P}} = \left[ {\begin{array}{*{20}{c}}
1&0&1&0&0&0\\
1&1&1&1&0&0\\
1&1&1&0&1&1\\
1&1&0&1&0&0
\end{array}} \right].
\end{align}

At the transmitter side, In contrast to the matrix design of SCMA in \eqref{SCMA}, where the number of users in each RB has to be the same, PDMA
allows a variable number of users to be mapped to each RB. For example, the first user in \eqref{PDMA} is seen to activate all the subcarriers for transmission,
while the sixth user only activates a single subcarrier. Another main difference  between PDMA and SCMA is that PDMA does not aim for achieving a constellation
shaping gain, which beneficially avoids the complex multi-dimensional constellation design of SCMA. At the receiver side, the MPA may also be adopted
for joint decoding, which is similar to LDS-CDMA and SCMA. Moreover, other MPA-SIC detection strategies or the turbo detection design of \cite{Ren2016PDMA}
can also be used for improving the decoding performance.


\begin{table*}[t!]\tiny
\caption{Comparison of the existing  NOMA solutions}
\begin{center}
\centering
\begin{tabular}{|l|l|l|l|l|l|l|l|l|}
\hline
\centering
 \textbf{Paradigm} &\textbf{Carrier} &\textbf{Spreading} & \textbf{Modulation} & \textbf{Flexibility}& \textbf{Complexity} & \textbf{Decoding} &\textbf{Key Advantages} & \textbf{Main Challenges}\\
\hline
\centering
PD-NOMA &  SC & Not used& Separated  & High & Low & SIC& High Spectral Efficiency & Error Propagation\\
\hline
\centering
IDMA & SC& Low-rate& Separated & Medium& Low & IMD & High Diversity Gain &  Interleaver Design\\
\hline
\centering
LDS-CDMA & SC& Sparse& Separated & High& Medium & MPA & CSI is not required &  Coding Redundancy\\
\hline
\centering
LPMA& SC& Lattice& Separated & Medium & Medium & SIC&Gain benefits PD and CD & Specific Channel Coding\\
\hline
\centering
LDS-OFDM& MC& Sparse& Integrated & Low & Medium & MPA & Wideband Signals Suitable& Coding Redundancy\\
\hline
\centering
SCMA& MC& Sparse& Integrated & Low  & High & MPA& Enhanced Diversity than LDS & Codebook Design\\
\hline
\centering
PDMA& MC& Sparse&Separated & Medium & High & MPA-SIC&Multi-dimension Diversity &Patterns
 Design \\
\hline
\end{tabular}
\end{center}
\label{table:NOMA variations}
\end{table*}

\subsection{Discussions and Outlook}
While we have provided a detailed introduction to the most popular power-domain NOMA variants,  there are also
other NOMA schemes, such as multiuser superposition transmission (MUST) scheme of \cite{MUST2016ICC}, the MUSA arrangement of~\cite{Yuan2016Multi}, etc.
Table \ref{table:NOMA variations} compares their key characteristics at a glance, where ``SC" refers to single-carrier and ``MC" refers to
multi-carrier schemes. Note that each form has its own advantages and disadvantages. For example, SCMA is an open-loop grant-free access scheme, which is suitable
for uplink transmission, and it relies both on sophisticated codebook design as well as on joint MPA detection. MUST is capable of enhancing the bandwidth efficiency
but it can only be used for downlink transmission. As a consequence, a unified NOMA framework is sought, which is capable of supporting the access
of numerous users in general scenarios. This is a promising future problem for researchers to contribute to.  Additionally, more research
contributions are also needed on the following aspects: 1) Optimal codebook/coding designs for SCMA and PDMA; 2) Hybrid MPA-SIC designs are sought
for low-complexity decoding; 3) The joint designs of  new modulation and MA schemes.

\section{Summary and Conclusions}
In this article, the recent literature of power-domain multiplexing aided NOMA proposed for 5G systems has been surveyed with an emphasis on
the following aspects: the basic principles
of NOMA, the amalgams of multiple antenna techniques and NOMA, the interplay of NOMA and cooperative communications, the resource control of NOMA, its co-existence with other key 5G techniques, and the implementation challenges and standadization. Apart from surveying the existing NOMA contributions, we have highlighted the key advantages of
NOMA itself as well as the inherent features of other techniques in Section II. Investigating the inherent integration of multiple-antenna aided
and cooperative techniques with NOMA is particularly important, since they are capable of providing extra spatial diversity, either with the aid of
centralized or distributed beamforming designs. It is worth pointing that when designing centralized beamformers, the most influential factor is
the appropriate ordering of the matrix based channels. Design guidelines were provided both for multiple-antenna aided and cooperative
NOMA in Sections III and IV, respectively.

Bearing in mind that NOMA  multiplexes users associated with different power levels with the aid of superposition coding techniques, the power
sharing among the users should be carefully optimized for each scenario.  For example, in practical scenarios, a laptop displaying
high-definition online video may share a RB with low-rate wireless sensors. Clearly, these applications have very different target rates,
and processing capabilities. More particularly, high-capability  laptops and smart phones can readily carry out SIC,
but this is quite the contrary for sensors. As such, intelligent resource control including PA and user scheduling should be investigated
by taking into account the recommendations listed in Section V. What is also worth emphasizing is that the proposed software-defined NOMA strategy
is capable of dealing with all the above-mentioned issues from a global perspective for the sake of holistically optimizing the performance of the entire
network.

The benefits of invoking NOMA in combination with the emerging large-scale MIMO, cooperative transmission, wireless power transfer, HetNets, etc were discussed.
 However, given the fact that
NOMA imposes extra intra-user interference on the system,
it brings about a range of open challenges, especially those related to sophisticated interference management. This problem would become more challenging
in CR, D2D, and HetNets scenarios. Indeed, the related research  of NOMA  relying on
sophisticated interference coordination is still in its infancy and hence has to be more deeply investigated, as recommended in Section VI.

The implementation issues of NOMA such as error propagation, channel estimation errors, security issues, etc. were discussed in Section VIII. Moreover, the standardization progress of NOMA was also discussed. It was noting that there are still  several gaps between research and implementation to fill with, which require more efforts on this field, as described in Section VIII.

Apart from power-domain NOMA, other NOMA schemes have also been advocated both by the industry and academia, as described in Section VIII. All NOMA
schemes discussed in this treatise share the same spirits, of non-orthogonal transmissions to enhance the attainable bandwidth efficiency and to
provide connectivity for numerous users within the limited number of RBs. Naturally, the problem of designing a unified NOMA framework
would be beneficial for supporting 5G scenarios.

\bibliographystyle{IEEEtran}
\bibliography{mybib_thesis}

\end{document}